\documentclass[12pt,preprint]{aastex}

\begin{document}
\shorttitle{The $Chandra$ view of the CG 12} \shortauthors{Getman
et al.}

\slugcomment{Submitted to the Astrophysical Journal 05/08/07}

\title{The Stellar Population And Origin Of The Mysterious High-Latitude Star Forming Cloud CG 12}

\author{Konstantin V.\ Getman\altaffilmark{1}, Eric D.\ Feigelson\altaffilmark{1}, Warrick A.\ Lawson\altaffilmark{2}, Patrick S. Broos\altaffilmark{1}, Gordon P.\ Garmire\altaffilmark{1}}

\altaffiltext{1}{Department of Astronomy \& Astrophysics, 525
Davey Laboratory, Pennsylvania State University, University Park
PA 16802}

\altaffiltext{2}{School of Physical, Environmental, and
Mathematical Sciences, University of New South Wales, Australian
Defence Force Academy, Canberra, ACT 2600, Australia}

\email{gkosta@astro.psu.edu}

\begin{abstract}

The mysterious high galactic latitude cometary globule (CG) CG~12
has been observed with the ACIS detector on board the $Chandra$
X-ray Observatory. We detect 128 X-ray sources, of which a half
are likely young stars formed within the globule's head. This new
population of $\ga 50$ T-Tauri stars and one new embedded
protostar is far larger than the previously reported few
intermediate-mass and two protostellar members of the cloud. Most
of the newly discovered stars have masses $0.2-0.7$~M$_\odot$, and
$9-15\%$ have $K$--band excesses from inner protoplanetary disks.
X-ray properties provide an independent distance estimate
consistent with CG~12's unusual location $\ga 200$~pc above the
Galactic plane. The star formation efficiency in CG~12 appears to
be $15-35$\%, far above that seen in other triggered molecular
globules. The median photometric age found for the T-Tauri
population is $\sim 4$~Myr with a large spread of $<1-20$~Myr and
ongoing star formation in the molecular cores. The stellar age and
spatial distributions are inconsistent with a simple radiation
driven implosion (RDI) model, and suggest either that CG~12 is an
atypically large shocked globule, or it has been subject to
several distinct episodes of triggering and ablation. We report a
previously unnoticed group of B-type stars northwest of CG~12
which may be the remnants of an OB association which produced
multiple supernova explosions that could have shocked and ablated
the cloud over a 15-30~Myr period. HD~120958 (B3e), the most
luminous member of the group, may be currently driving an RDI
shock into the CG~12 cloud.

\end{abstract}


\section{Introduction \label{introduction_section}}

Star formation in the Galaxy occurs on all scales from a small
globule collapsing into a single star with planetary system to a
giant molecular cloud complex forming rich stellar clusters
\citep{Elmegreen00}. Most stars form in gravitationally unbound
groups, and many of these are relatively sparse groups with
$10-100$ members and initial surface density $\la 100$
stars/pc$^2$ \citep{Clarke00}. Unbound sparse clusters are
difficult to study because their members quickly disperse and are
not easily discriminated from the older Galactic field population.
X-ray selection is effective here: median X-ray luminosities of
pre-main sequence stars are 30 (5) times higher than ZAMS
Pleiades-age stars, and 500 (500) times higher than solar
neighborhood stars for $M \simeq 1$ (0.3) M$_\odot$ stars
\citep{Preibisch05a}. X-ray surveys are also effective in
identifying young stars which have lost their protoplanetary
disks, and thus effectively complement infrared surveys that
preferentially select for disk emission \citep{Feigelson07}. For
example, {\it ROSAT} studies were responsible for locating most
known dispersed $\sim 10$ Myr old stars associated with the TW Hya
and $\eta$ Cha associations \citep[e.g.][]{Webb99,Mamajek00},
while {\it XMM-Newton} studies are helping to identify $\sim
5$~Myr old members of the Upper Scorpius association
\citep{Argiroffi06}. $Chandra$ exposures resolve substantial
pre-main sequence (PMS) populations embedded in cloud cores
\citep[e.g., NGC~1333,][]{Getman02}, in rich clusters \citep[e.g.,
Orion Nebula cluster,][]{Getman05a}, and in sparse gravitationally
bound multiple systems \citep{Feigelson03}.

Some progress is being made towards characterizing the population,
dynamics, and particularly the protoplanetary disk evolution of
stars born in sparse clusters. \citet{Megeath05} employed
$Spitzer$ IRAC photometry to detect a $40\%$ disk fraction around
late-type members of the 8~Myr old $\eta$ Cha cluster at
$d=100$~pc. This disk fraction is higher than expected from
studies of richer clusters where a 6~Myr maximum lifetime for
inner disks was derived by \citet{Haisch01}. This argues against a
single dissipation timescale for all environments and all disk
radii. $Spitzer$ IRS spectroscopy of $\eta$ Cha stars by
\citet{Bouwman06} show that the dissipation timescale may be a
strong function of binarity, with disks lasting longer around
single stars.

We report here a new X-ray selected sample of young stars to
further these studies. The target is cometary globule CG~12 (= MBM
112) \citep{Hawarden76}, a relatively poorly studied small star
forming region at high galactic latitude
$(l,b)=(316^\circ,21^\circ)$, which is producing a sparse unbound
cluster of young stars. Shown in Figure~\ref{fig_dss_combined},
the head of CG~12 is the reflection nebula NGC~5367 (= GN 13.54.7)
illuminated by the double B4 + B7 star Herschel~4636 (= CD
-39$^\circ$8581) \citep{Herschel1847}. Signs of ongoing star
formation include CO molecular cores
\citep{vanTill75,White93,Yonekura99,Haikala06}, several IRAS
sources \citep{White93,Bourke95a}, a bipolar molecular outflow
\citep{White93}, and three stars with K-band excesses
\citep{Santos98}. Optical photometry and spectroscopy of the
brightest stars in the region reveal three B type (B4, B7, B8) and
two A type (A2 + A4) candidate members
\citep{vanTill75,Williams77,Maheswar04}. The most recent mass
estimate for the total of all three molecular cloud cores
(CG~12SW, CG~12S, and CG~12N) found in CG~12, ranges from $100$ to
$250$~M$_{\odot}$ with the peak column density through the densest
core, CG~12N, of $\sim 2.2 \times 10^{22}$~cm$^{-2}$ ($A_V \sim
13$~mag) \citep{Haikala06}.  Due to its high Galactic latitude,
contamination by field stars and obscuration by foreground
interstellar matter is much lower than for most star forming
regions.

Compact CGs, along with other Bok globules, are interpreted as the
last vestiges of the molecular cloud that gave birth to OB stars
\citep{Bok47,Hawarden76}. CGs are distributed widely across the
Milky Way but are rarely found at high latitude \citep{Bourke95a}.
Most lie within giant HII regions such as IC~2944, the Rosette and
Carina nebulae, and most prominently in the huge expanding Gum
Nebulae surrounding the Vela OB2 association. CGs are modeled as
externally illuminated, photoevaporated and ablated into elongated
head-tail morphologies by ultraviolet radiation of OB stars
\citep{Reipurth83}.  It is likely that pressure from the
ionization shock front at the surface propagates through a globule
and overcomes the magnetic, turbulent and thermal pressure that
supports it against collapse, thereby triggering star formation.
This process of radiation-driven implosion (RDI) in molecular
globules was originally studied through analytical models for
spherical clouds \citep{Bertoldi89} and 2-dimensional
hydrodynamical calculations \citep{Lefloch94,Lefloch95}. Recent
3-dimensional calculations of RDI processes in molecular globules,
including self-gravity, heating, and cooling, are presented by
\citet{Kessel03} and \citet{Miao06}.  The characteristic timescale
for producing cometary morphologies and inducing gravitational
collapse is $0.1-0.5$~Myr.

Generally, the stellar populations of these isolated triggered
star formation regions are poorly known. Some globules show clear
evidence for embedded young stars, such as {\it IRAS} sources,
while others do not. H$\alpha$ and $JHK$ surveys of bright-rimmed
globules containing embedded IRAS sources revealed several cases
with small clusters of T~Tauri and $K$-band excess stars within
and in front of the bright rim \citep{Sugitani95, Ogura02,
Thompson04}. Molecular outflows and masers are found in several
globule cores with IRAS sources \citep{deVries02, Urquhart06}.
Directly supporting the RDI triggering mechanism, age gradients in
the stellar population (where the youngest stars are embedded and
older stars are aligned toward the ionizing sources) have been
recently reported in irradiated molecular globules within the
large IC~1848 and IC~1396 HII regions
\citep{Matsuyanagi06,Ogura06,Getman07}. For example, our
$Chandra$/$Spitzer$ study of IC~1396N revealed a spatial-age
gradient in the globule from Class~III to Class~I/0 among 25 young
objects \citep{Getman07}.

The existence of any star forming dark cloud at Galactic location
$(l,b)=(316^\circ,21^\circ)$ is unusual, and the cometary
structure of CG~12 in such an isolated environment is even more
mysterious. There are no reported luminous OB-type stars at even
higher latitudes to produce the head-tail morphology with its tail
pointing towards (not away from) the Galactic plane.  The origin
of CG~12 is thus uncertain. \citet{Williams77} suggested that a
high latitude supernova explosion ($l \simeq 320^{\circ}, b \simeq
30^{\circ}$) could have ablated the cloud and triggered star
formation. We develop these ideas here and identify a specific
stellar group which may be responsible for compressing and
ablating the CG~12 cloud.

The CG~12 population is likely to be far larger than that of
previously studied molecular globules. The presence of five
intermediate-mass ($2<M<7$ M$_\odot$) B4-A4 stars
\citep{Williams77,Maheswar04} demands the presence of $\sim 30$
stars with masses $0.5<M<2$ M$_\odot$ and $\sim 50$ stars with
masses $0.2<M<0.5$ M$_\odot$, assuming a Galactic field Initial
Mass Function \citep{Kroupa02}. Thus, roughly a hundred cloud
members are expected to be present. A fraction of these stars have
already been detected in X-rays, although the result has not been
published.  An archived 9 ks $ROSAT$-PSPC image (obtained to study
shadowing of the soft X-ray background by CG~12) harbors $\sim 30$
soft X-ray sources within the globule's head and in the arc around
the globule (Figure \ref{fig_rosat}). Most of those are clearly
not background sources, as the remainder of the $ROSAT$ field is
nearly empty. CG~12 thus appears to be the most active star
forming center among the known cometary globules, and should
provide a valuable laboratory for testing theories of triggered
star formation.

This paper begins a study of the CG~12 stellar population based on
an X-ray survey designed to address the issues outlined above. We
discuss the population, spatial and age distribution, dynamics,
and star formation history. We suggest how a CG morphology arose
at this atypical isolated location. Our work is based on a mosaic
of four images obtained with the $Chandra$ X-ray Observatory
combined with published optical, infrared and radio data. With its
excellent mirrors and detectors, $Chandra$ achieves an
order-of-magnitude improvement in sensitivity and spatial
resolution compared to $ROSAT$.  Our planned new optical and
$Spitzer$-IRAC studies of the CG~12 stellar population will be
described in subsequent papers.

The $Chandra$ observations of CG~12 and source list are described
in \S \ref{observation_section}, optical and near-infrared (ONIR)
counterparts to $Chandra$ sources are presented in \S
\ref{counterparts_section}, and evaluation of X-ray contaminants
unrelated to the cloud is provided in \S
\ref{contamination_section}. The X-ray and ONIR properties of most
of the discovered T-Tauri stars with a preliminary analysis of
their disks and kinematics are reported in \S
\ref{sepctral_analysis_section} and \ref{ONIR_photometry_section}.
Discussion of the distance to CG~12 is presented in \S
\ref{distance_section}. The inferred global properties of the
CG~12 stellar population, including its age distribution, star
formation efficiency, and spatial structure appear in \S
\ref{stellar_population_section}. We end in \S
\ref{discussion_section} with a discussion of the new
observational findings and their implications for the history of
star formation and origin of CG~12.

\section{The {\it Chandra} Observations and Analysis
\label{observation_section}}

\subsection{Observation and Data Reduction \label{data_reduction_section}}

\begin{figure}
\centering
\includegraphics[angle=0.,width=6.0in]{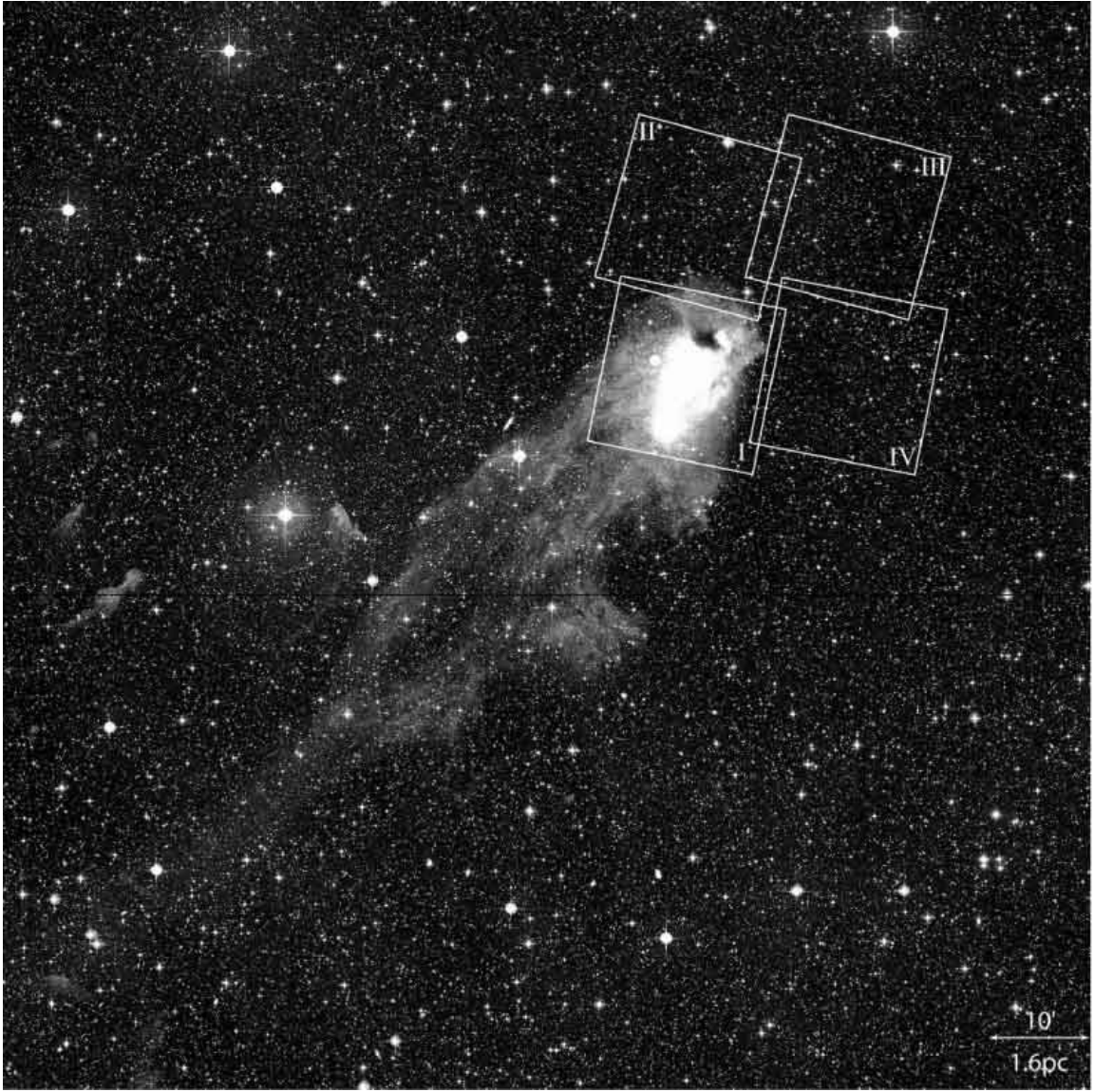}
\caption{An optical image covering $\sim 1.8\arcdeg \times
1.8\arcdeg$ of the CG~12 neighborhood from the Digitized Sky
Survey (DSS). The four $Chandra$ ACIS-I fields, $17\arcmin \times
17\arcmin$ each, are outlined and labelled.  A dark molecular
cloud lies in the head of the globule, and is superposed by three
nebulous reflection nebulae: NGC~5367 (=GN 13.54.7), GN~13.54.9,
and B77 146. \label{fig_dss_combined}}
\end{figure}

\begin{figure}
\centering
\includegraphics[angle=0.,width=6.0in]{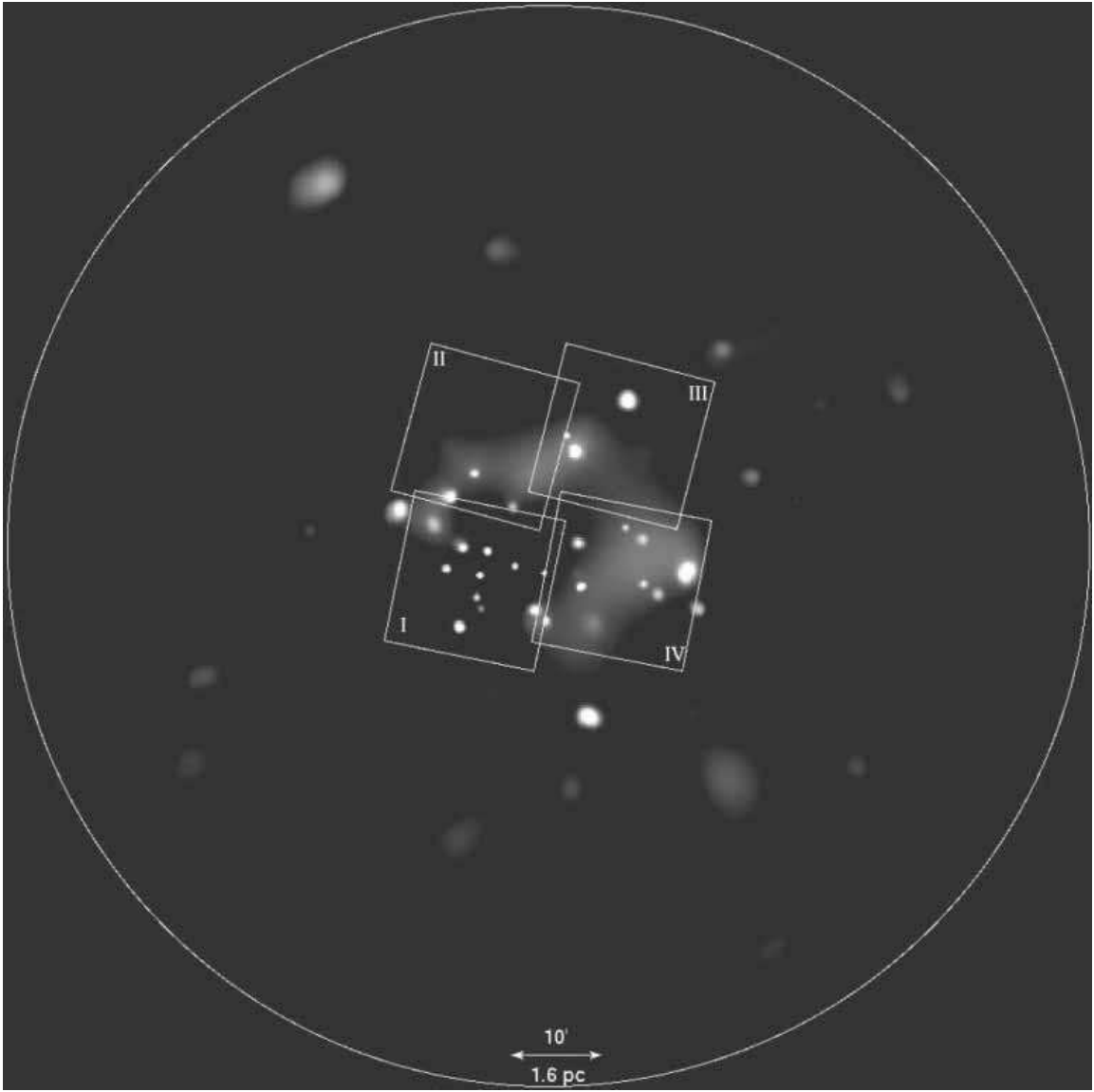}
\caption{Adaptively smoothed $ROSAT$-PSPC image of the CG~12
neighborhood with logarithmic gray scales. The four ACIS-I
$Chandra$ fields are outlined and labelled. The $\sim 2\arcdeg$ in
diameter $ROSAT$-PSPC field of view is outlined by the white
circle. About 30 X-ray point sources are seen on the image.
\label{fig_rosat}}
\end{figure}

The current project combines four X-ray observations of the
globule, one primary field (I in Fig. \ref{fig_dss_combined}) with
$\sim 31$~ks exposure directed at the globule's core and three
secondary fields (II, III, and IV in Fig. \ref{fig_dss_combined})
with $\sim 3$~ks exposures positioned contiguously to the north
and west of the core. The primary pointing is intended to detect
the population of pre-main sequence stars forming in the molecular
head of the globule. The secondary pointings are designed to
locate an older population of stars expected if the present cloud
is only the ablated remnant of a larger cloud that experienced
sequential star formation triggering events, similar to the
sequence of stars found in our $Chandra$ study of IC~1396N
\citep{Getman07}.

The observations were obtained with the ACIS camera
\citep{Garmire03} on-board {\it Chandra} \citep{Weisskopf02} as
described in Table 1. For each observation we consider here only
results arising from the imaging array (ACIS-I) of four abutted
$1024 \times 1024$ pixel front-side illuminated charge-coupled
devices (CCDs) covering about $17\arcmin \times 17 \arcmin$ on the
sky. The S2 and S3 detectors in the spectroscopic array (ACIS-S)
were also operational, but as the telescope point spread function
(PSF) is considerably degraded far off-axis, the S2-S3 data are
omitted from the analysis. All four observations were performed
during a quiet phase of solar activity; no background flares due
to particle contamination were encountered.

Data reduction follows procedures similar to those described in
detail by \citet[][Appendix B]{Townsley03} and \citet{Getman05a}.
Briefly, the data are partially corrected for CCD charge transfer
inefficiency \citep{Townsley02}, selected by event ``grade" (only
ASCA grades 0, 2, 3, 4, and 6 are accepted), ``status", and
``good-time interval" filters, trimmed of background events
outside of the $0.5-8.0$ keV band, and cleaned of hot pixel
columns with energies of $< 0.7$ keV left by the standard
processing. The slight PSF broadening from the $Chandra$ X-ray
Center's (CXC's) software randomization of positions is removed.
Based on several dozen matches between bright $Chandra$ and 2MASS
point sources, we applied a small astrometric correction to source
positions to place the absolute astrometry to the {\it Hipparcos}
reference frame. These and later procedures were performed with
$CIAO$ software package 3.3, $CALDB$ 3.2.1, $HEASOFT$ 6.1.1,
\anchor{http://www.astro.psu.edu/users/townsley/cti/}{CTI
corrector} version 1.44, and the
\anchor{http://www.astro.psu.edu/xray/docs/TARA/ae\_users\_guide.html}{{\it
ACIS Extract}} package version 3.94. The latter two tools were
developed at Penn State and are publicly
available\footnote{Descriptions and codes for CTI correction and
{\it ACIS Extract} can be found at
\url{http://www.astro.psu.edu/users/townsley/cti/} and
\url{http://www.astro.psu.edu/xray/docs/TARA/ae\_users\_guide.html},
respectively.}.

\subsection{X-ray Point Source Catalog \label{srclist_section}}

\begin{figure}
\centering
\includegraphics[angle=0.,width=6.0in]{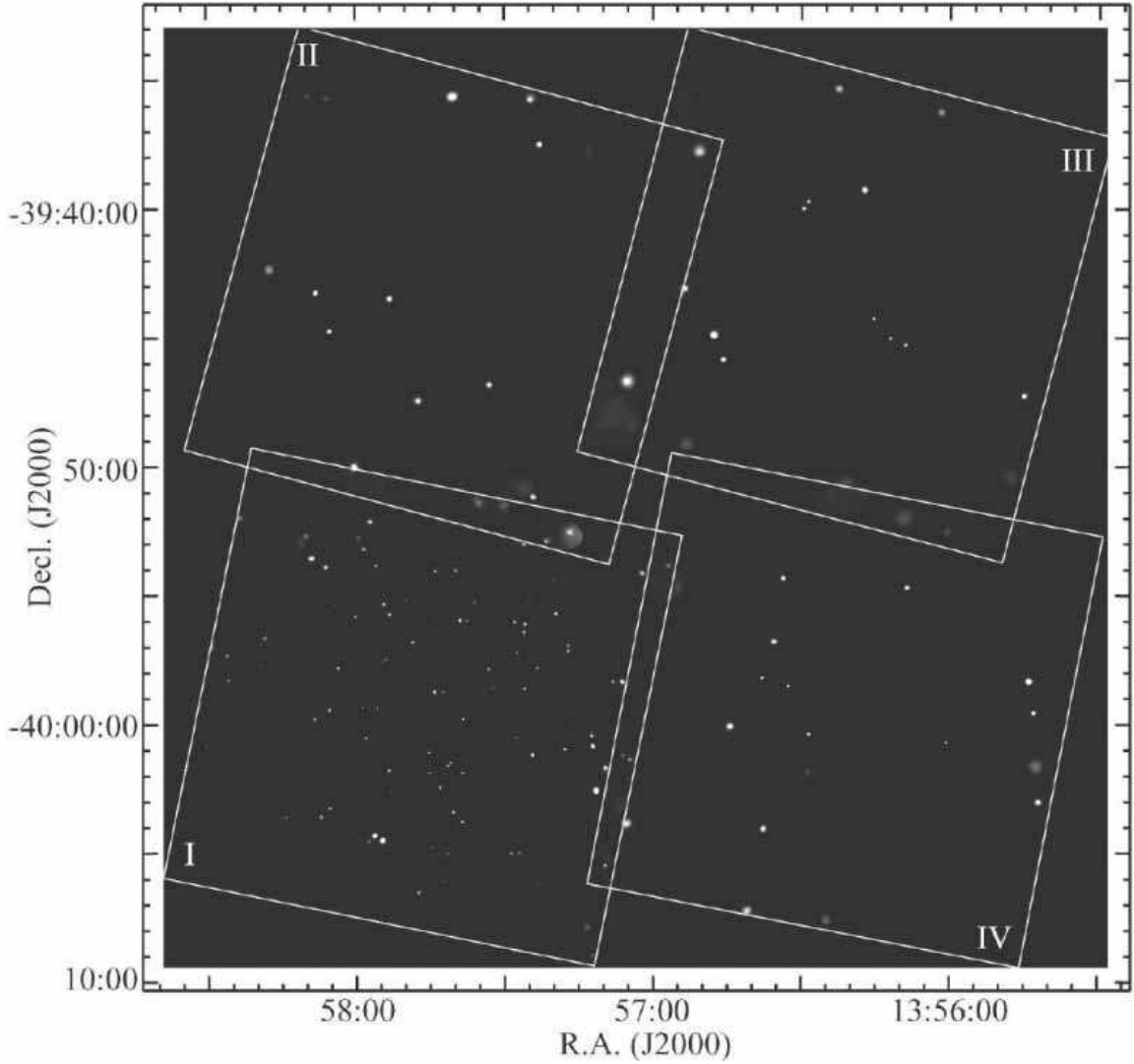}
\caption{Mosaicked and adaptively smoothed $Chandra$ ACIS-I image
of the CG~12 neighborhood. Smoothing has been performed in the
full $0.5-8.0$ keV band at the 2.5$\sigma$ level, and gray scales
are logarithmic. The four ACIS-I fields are outlined and labelled;
most of the 128 X-ray sources can be seen.
\label{fig_csmooth_combined}}
\end{figure}

Figure \ref{fig_csmooth_combined} shows the resulting image of the
ACIS-I field after adaptive smoothing with the CIAO tool
$csmooth$. More than 100 point sources are easily discerned.
Source searching was performed with data images and exposure maps
constructed at three spatial resolutions (0.5, 1.0, and
$1.4\arcsec$ per pixel) using the $CIAO$ {\it wavdetect} tool. We
run {\it wavdetect} with a low threshold $P = 10^{-5}$ which is
highly sensitive but permits false detections at this point in the
analysis. This is followed by visual examination to locate other
candidate sources, mainly close doubles and candidate sources near
the detection threshold. Using
\anchor{http://www.astro.psu.edu/xray/docs/TARA/ae\_users\_guide.html}{{\it
ACIS Extract}}, photons are extracted within polygonal contours of
$\sim 90\%$ encircled energy using position-dependent models of
the PSF.

Background is measured locally in source-free regions. Due to the
very low, spatially invariant, ACIS-I background in the $Chandra$
observations of CG~12 there is a one-to-one correspondence between
a source's significance and net counts \citep[a similar
correspondence is seen in the $Chandra$ data of IC~1396 globule,
see Figure 2 of][]{Getman07}.  This uniform background is much
simpler than in the $Chandra$ Orion Ultradeep Project (COUP)
observation of the Orion Nebula Cluster where source crowding,
readout trails and PSF wings from bright sources complicate the
analysis \citep{Getman05a}. The source significance calculated by
\anchor{http://www.astro.psu.edu/xray/docs/TARA/ae\_users\_guide.html}{{\it
ACIS Extract}} is the ratio of the source's net
(background-subtracted) counts to the uncertainty on that
quantity\footnote{The uncertainty on the net counts is estimated
by calculating errors \citep[][eqs. 7 and 12]{Gehrels86} on total
observed and background counts and then propagating \citep[][eq.
1.31]{Lyons91} those errors to net counts. More detailed
information on the algorithm for calculating source significance
can be found in the section ``Algorithms'', subsection ``Broad
Band Photometry'', of the Users Guide for {\it ACIS Extract} at
\url{http://www.astro.psu.edu/xray/docs/TARA/ae\_users\_guide.html}.}:
$Signif = NetFull/\sigma(NetFull)$. Following the procedure of
\citet{Getman07}, the list of candidate sources is trimmed to omit
sources with fewer than $\sim 5$ estimated source net counts,
$NetFull/PSFfrac \lesssim 4.5$.  In the case of CG~12 observations
the above criterion is equivalent to accepting sources with a
source significance of $Signif \ga 1.1$. The resulting catalog of
128 sources is listed in Table \ref{tbl_bsp}.

The
\anchor{http://www.astro.psu.edu/xray/docs/TARA/ae\_users\_guide.html}{{\it
ACIS Extract}} package estimates (often using $CIAO$ tools) a
variety of source characteristics including celestial position,
off-axis angle, net and background counts within the PSF-based
extraction area, source significance assuming Poisson statistics,
effective exposure (corrected for telescope vignetting and
satellite dithering), median energy after background
subtraction\footnote{The derivation of median energy of
background-subtracted events is given in \citet[][\S
3.1]{Townsley06}.}, a variability indicator derived from the
one-sided Kolmogorov-Smirnov statistic, and occasional anomalies
related to chip gap or field edge positions. These are all
reported in Table \ref{tbl_bsp}; see \citet{Getman05a} for a full
description of these quantities.

Figure \ref{fig_combined_rgb} summarizes extracted properties of
the 128 $Chandra$ point sources in four dimensions: source
position on the sky, median energy of the source photons, and
count rate. The three known B-type members of the region are
detected by $Chandra$ (\S \ref{stellar_census_section}). These are
$Chandra$ sources I$_{-}57$, I$_{-}58$, and I$_{-}74$. Most of the
unabsorbed X-ray point sources (red circles in Figure
\ref{fig_combined_rgb}) are likely previously unreported T-Tauri
stars formed in the cloud (\S \ref{contamination_section},
\ref{ONIR_photometry_section}, and \ref{ages_unobscured_section}).
Several bright hard X-ray sources (large blue circles) coincide
with the C$^{18}$O molecular cores of \citet{Haikala06}: I$_{-}45$
and I$_{-}46$ lie within CG~12N, and I$_{-}54$ lies within CG~12S.
These are probably embedded protostars  (\S
\ref{ONIR_photometry_section} and \ref{embedded_ages_section}).
Most of the remaining X-ray sources (small cyan and blue circles)
have no stellar counterparts; many of these should be
extragalactic contaminants (\S \ref{contamination_section}).

\subsection{X-ray spectral analysis
\label{sepctral_analysis_section}}

For $Chandra$ sources with $>20$ net counts, we performed spectral
analysis with the $XSPEC$ spectral fitting package version 12.2
\citep{Arnaud96}. The unbinned source and background spectra were
fitted with one-temperature APEC plasma emission models
\citep{Smith01} using the maximum likelihood method which is
believed to be well suited to low-count sources \citep{Cash79}. We
assumed 0.3 times solar elemental abundances previously suggested
as typical for young stellar objects in other star forming regions
\citep{Imanishi01,Feigelson02}. Solar abundances were taken from
\citet{Anders89}. X-ray absorption was modeled using the atomic
cross sections of \citet{Morrison83}.

Best-fit plasma parameters inferred from our spectral fits and
derived broadband luminosities are tabulated in Table
\ref{tbl_thermal_spectroscopy}.  See table notes for further
details. The derived parameters are generally consistent with the
correlation between the absorbing column density and photometric
source median energy ($MedE$) found for Orion stars in our COUP
study \citep[see Figure~8 in][]{Feigelson05}.   In several cases
where the source is not truly associated with a CG~12 star (\S
\ref{contamination_section}),  source luminosities reported in
Table \ref{tbl_thermal_spectroscopy} are obviously wrong. For
instance, hard X-ray source I$_{-}2$ outside of the molecular
cloud is a good extragalactic candidate, while an extremely soft
X-ray source I$_{-}81$ is likely to be a foreground star (\S
\ref{contamination_section}).

\section{ONIR counterparts \label{counterparts_section}}

$Chandra$ source positions were compared with source positions
from three ONIR catalogs. The Naval Observatory Merged Astrometric
Dataset \citep[NOMAD;][]{Zacharias04a} contains astrometric and
photometric data for over 1 billion stars derived from the
Hipparcos, Tycho-2, UCAC2, and USNO-B catalogs.  A list of stars
around CG~12 with $BVRI$ photometry is provided by
\citet{Maheswar04}.  The Two Micron All-Sky Survey
\citep[2MASS;][]{Cutri03} catalog gives sources down to $K \sim
15.5$~mag. $Chandra$ sources were considered to have stellar
counterparts when the positional coincidences were better than
$1\arcsec$ within $\sim 3.5\arcmin$ of ACIS-I field center, and
$2\arcsec$ in the outer regions of each ACIS-I field where
$Chandra$ PSF deteriorates. $Chandra$-2MASS ($Chandra$-NOMAD)
median offsets are better than $0.2\arcsec$ ($0.3\arcsec$) at the
central part of the field and $0.4\arcsec$ ($0.5\arcsec$) at
larger off-axis angles, which are similar to those found in the
COUP observation \citep{Getman05a}. We find that NOMAD is slightly
more sensitive than 2MASS catalog for $Chandra$ counterparts: 55
(out of 128) X-ray sources have optical counterparts while 47 have
2MASS counterparts. All five with optical but without 2MASS
counterparts are faint with $R \sim 19$~mag sources; some of these
may be extragalactic.

Individual ONIR counterparts for $Chandra$ sources are listed in
Table \ref{tbl_cntrpts} along with their optical $VR$ and NIR
$JHK_s$ magnitudes and proper motions.  See table notes for
details. When an optical counterpart from \citet{Maheswar04} is
present, its more reliable CCD photometry supersedes the
photographic photometry from the NOMAD catalog.

\begin{figure}
\centering
\includegraphics[angle=0.,width=6.0in]{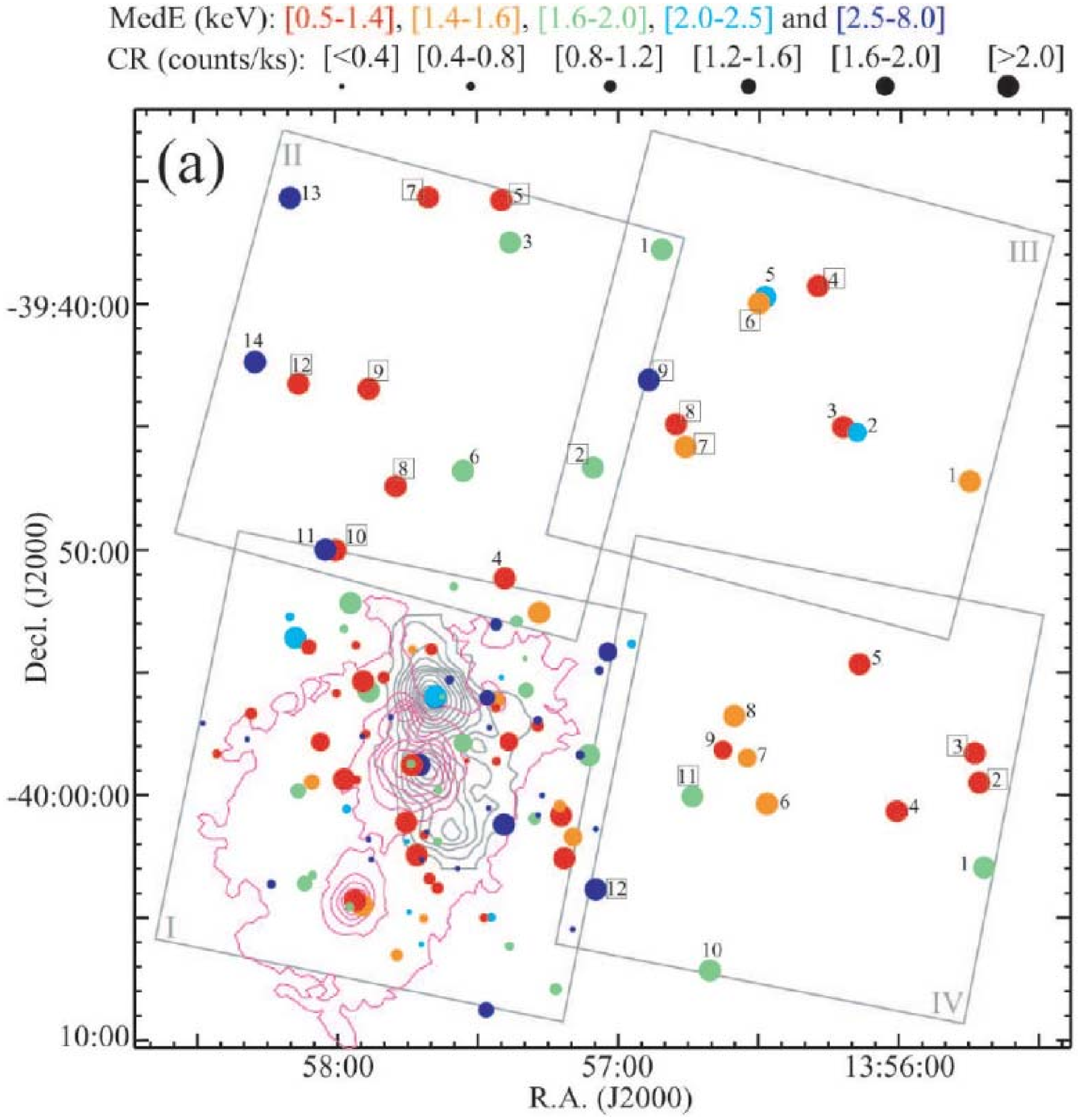}
\caption{\footnotesize Schematic representation of the X-ray
source properties in four dimensions:  R.A., Decl., median energy,
and count rate. The code for symbol colors is given at the top.
Symbol size is scaled to source count rate so that weakest sources
with ${\rm CR}<0.4$ cts s$^{-1}$ are represented by smallest
circles and ${\rm CR}>2$ cts s$^{-1}$ by the largest circles.  The
magenta contour outlines the optical nebulae in the head of the
globule seen in Figure \ref{fig_dss_combined}. The grey contour
marks the C$^{18}$O(1-0) emission measured by \citet{Haikala06}
with contour intervals of $0.3$~K~km/s which roughly corresponds
to 1~mag of visual absorption through the cloud. X-ray sources
with ONIR counterparts are outlined by $\square$; most of the
others are extragalactic contaminants. Panel a: The four $Chandra$
fields. Panel b: Close-up view of the primary field.  Reflection
nebulae are labelled in magenta
\citep{Bernes77,Neckel90,Magakian03} and molecular cores are
labelled in grey \citep{Haikala06}. \label{fig_combined_rgb}}
\end{figure}

\begin{figure}
\centering
\includegraphics[angle=0.,width=6.0in]{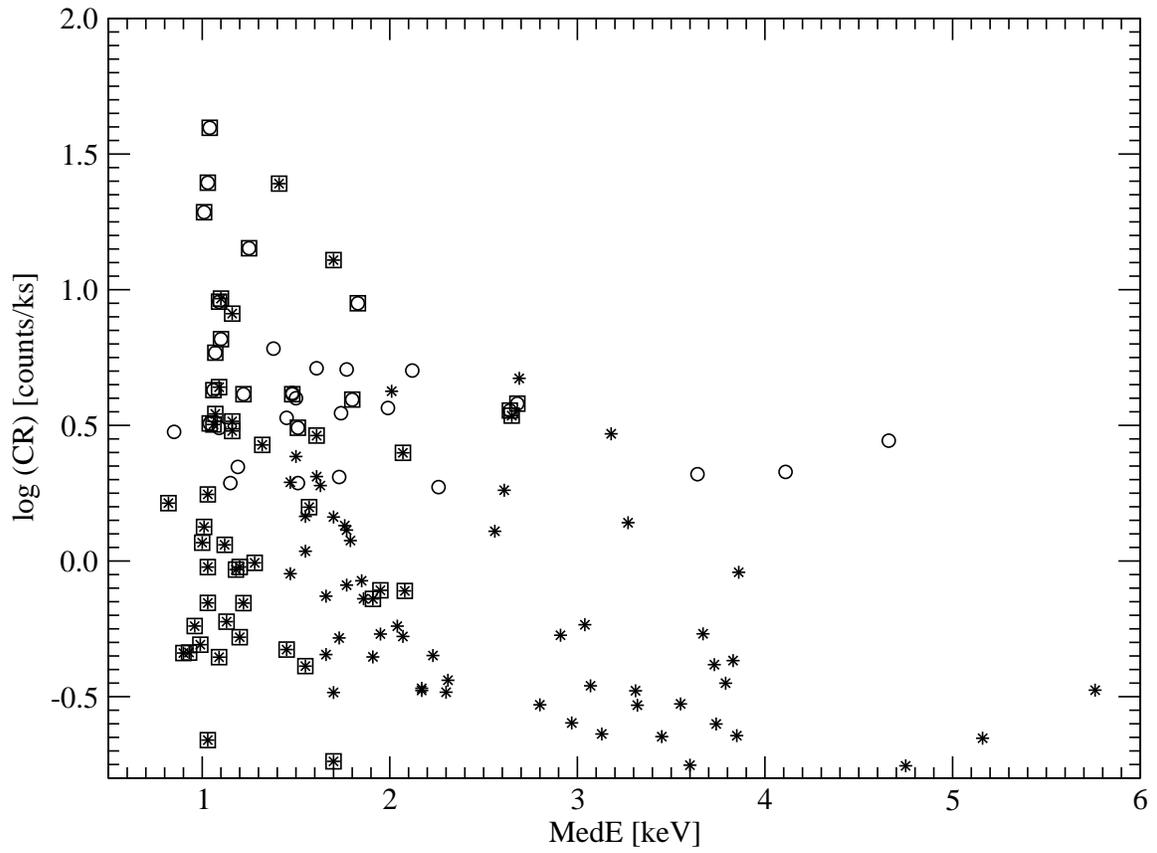}
\caption{X-ray count rate versus X-ray median energy for the 128
$Chandra$ sources. Field I sources are represented by star symbols
and Field II$-$IV sources are shown as circles.  X-ray sources
with optical counterparts are outlined by $\square$.
\label{fig_cr_vs_mede}}
\end{figure}

We established in our COUP study of Orion stars that the X-ray
median energy, $MedE$, of a PMS star is an excellent surrogate for
its line-of-sight obscuration, $N_H$ \citep{Feigelson05}. Figure
\ref{fig_cr_vs_mede} examines the relationship between the
presence of an optical counterpart, the X-ray count rate, and
X-ray median energy derived from the X-ray photometry (Table
\ref{tbl_bsp}). The presence of optical counterparts correlates
well with low median energy ($MedE < 2$~keV corresponding to $\log
N_H < 22.0$~cm$^{-2}$), but shows little relation to the count
rate. From this figure, we roughly partition the $Chandra$ sources
into three groups which are discriminated by symbol color in
Figure \ref{fig_combined_rgb}:
\begin{enumerate}

\item Group 1 consists of the 42 unobscured or lightly
obscured sources with ${\rm MedE} \leqslant 1.4$~keV (red circles
in Figure \ref{fig_combined_rgb}) of which 88\% are identified
with ONIR sources down to $R \sim 19$~mag.  All 27 of these
sources from field I have optical counterparts.  In fields II-IV,
5 out of 15 sources lack optical counterparts (II$_{-}4$,
III$_{-}3$, IV$_{-}4$, IV$_{-}5$ and IV$_{-}9$).

\item Group 2 consists of 54 lightly-moderately and some
heavily obscured $Chandra$ sources with $1.4 < MedE \leqslant
2.5$~keV (orange, green, and cyan objects in Figure
\ref{fig_combined_rgb}).  Only 28\% of these sources are
identified with ONIR counterparts.

\item Group 3 consists of the 32 heavily obscured sources or
intrinsically hard X-ray sources with $MedE > 2.5$~keV (blue
objects in Figure \ref{fig_combined_rgb}).  Only 9\% of these
sources have been identified in ONIR bands.

\end{enumerate}

\begin{figure}
\centering
\includegraphics[angle=0.,width=6.0in]{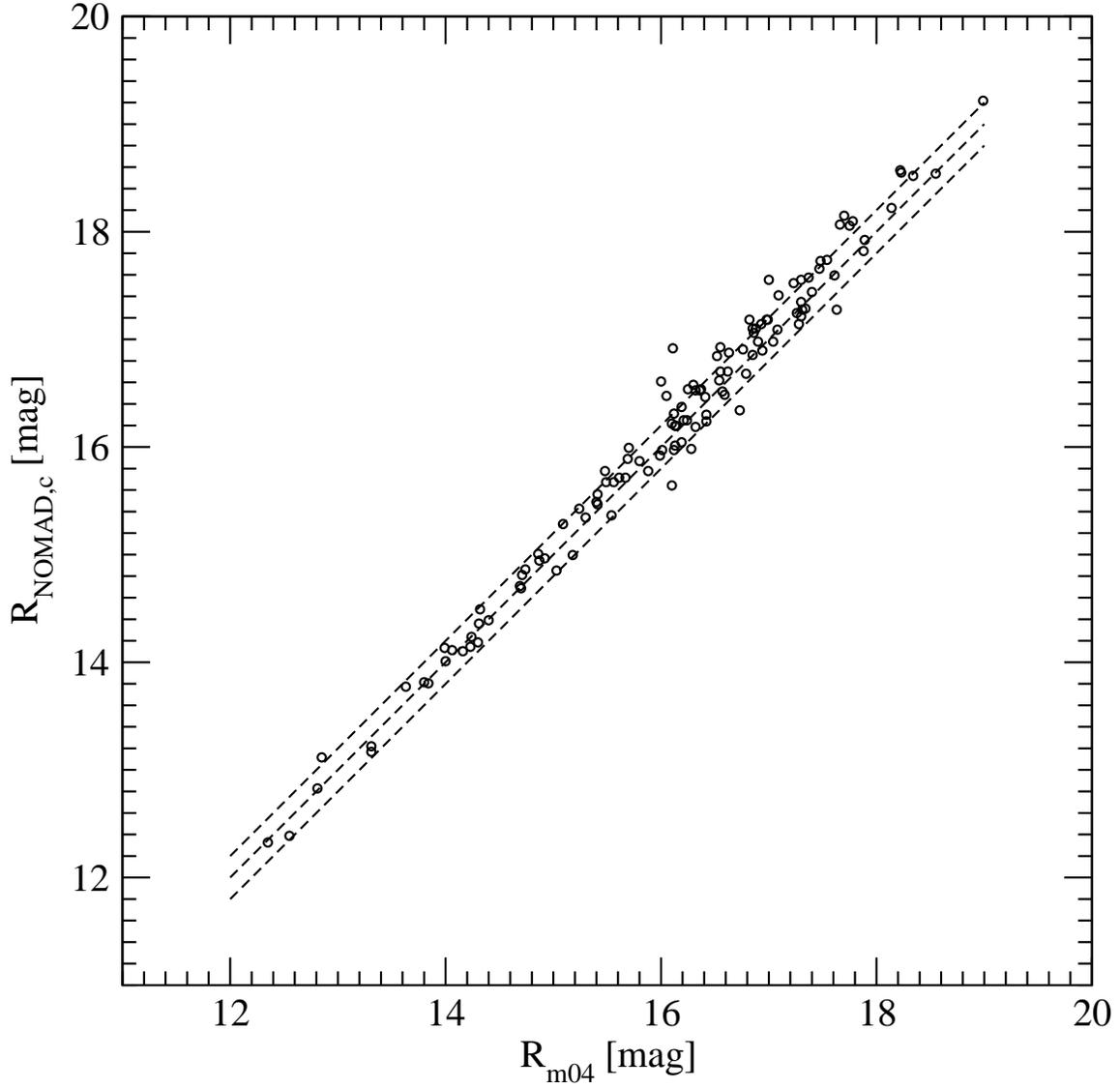}
\caption{Comparison of photographic $R$-band magnitudes (after
linear transformation) with CCD magnitudes for $120$ control
optical sources lying in the CG~12 field. Dashed lines show
$R_{NOMAD,c} - R_{m04}$ values of -0.2, 0, and
0.2.\label{fig_mah_vs_usno}}
\end{figure}

To facilitate later photometric study, we have calibrated the
photographic $R$-band magnitudes obtained from the NOMAD catalog
to the more accurate CCD magnitudes of \citet{Maheswar04}.  For
stars in both samples, we find that USNO-B $R$-band magnitudes
from NOMAD catalog are systematically overestimated by $\sim
0.5$~mag and can be mostly corrected by the linear correction
$R_{NOMAD,c} = 1.03 \times R_{NOMAD} - 0.92$. Figure
\ref{fig_mah_vs_usno} compares the corrected NOMAD and CCD
magnitudes.  From the figure, we see that the NOMAD values are
about 0.2~mag too faint for $R \ga 16.2$~mag. With these
additional corrections, the formal $1\sigma$ errors for $R$-band
magnitudes are better than $\sim 0.2$~mag for all $Chandra$
counterparts with $R > 12$~mag.

\section{Sources unrelated to the cloud}
\label{contamination_section}

Unlike in the highly concentrated ONC cluster in the COUP
observation, a significant fraction of X-ray sources in $Chandra$
observations of sparse clusters are unrelated to the star forming
region. To quantify this contamination, we performed detailed
Monte-Carlo simulations of Galactic field and extragalactic
background contamination populations as described in
\citet{Getman05b,Getman06}.

The Galactic field contamination is estimated from simulations of
Galactic stellar populations by \citet{Robin03}, henceforth the
Besan\c{c}on model\footnote{Simulations are provided at
\url{http://bison.obs-besancon.fr/modele/}.}.  We convolve the
simulated stellar population in a $Chandra$ field at $(l,b) =
(316.5, +21.1)$ with the X-ray luminosity functions of normal
Galactic disk stars in the solar neighborhood measured from
$ROSAT$ surveys \citep{Schmitt95, Schmitt97, Hunsch99} and then
apply the observational limit of our $Chandra$ observations. The
Besan\c{c}on model predicts $\sim 100$ foreground stars (74\% M,
10\% K, 10\% G, and 6\% F dwarfs) mostly at distances of
$250-550$~pc with one third having ages $<3$~Gyr. It predicts
$\sim 2700$ background stars down to $V < 22$~mag (18\% M, 34\% K,
28\% G, 18\% F dwarfs, and 2\% giants) mostly at distances of
$0.6-10$~kpc with $>3$~Gyr ages.

Convolving these distributions with the $ROSAT$ luminosity
functions, and including the effect of spatially-dependent
absorption derived from the C$^{18}$O column map of
\citet{Haikala06} (assuming a correspondence of $0.3$~K~km/s to
$A_V \sim 1$~mag), we obtain a predicted X-ray source
contamination in the $Chandra$ field~I of $\sim 4$~M and $\sim
1$~G-F dwarfs at distances around 300~pc. Source I$_{-}81$ is very
likely one of these contaminants; it has a very soft X-ray
spectrum, and has independently been suggested to be a G8$-$K5
field star unrelated to CG~12 \citep[WBLH \#4
in][]{Marraco78,Maheswar04}. No more than 1 foreground star of a
spectral type G or F is expected in each of the secondary
$Chandra$ fields (II, III, and IV). The simulations predict that
$\la 2$ background giants will be present in field I, and no
background dwarfs will be detected in any field.

Contamination by extragalactic X-ray sources will be more
important.  This is evaluated by applying the instrumental
background and obscuration effects of the molecular cloud to
simulated extragalactic X-ray source populations drawn from the
hard band X-ray background $\log N - \log S$ distribution of
\citet{Moretti03}.  Power-law spectra are assumed consistent with
flux-dependencies obtained by \citet{Brandt01}.

For each of the secondary $Chandra$ fields, the simulations
predict from $4-11$ extragalactic contaminants. The best AGN
candidates in our secondary fields are those without ONIR
counterparts; we see 19 such sources (Table \ref{tbl_cntrpts})
consistent with the simulations. Most have hard band observed
X-ray fluxes $\lesssim 10^{-13}$~ergs~cm$^{-2}$~s$^{-1}$ and
$R$-band magnitudes $> 19$~mag, consistent with measured
extragalactic X-ray flux-$R$ magnitude dependencies
\citep{Barger03,Steffen04}.

The simulations for the primary $Chandra$ field predict $\sim
40-50$ extragalactic contaminants: around half of the contaminants
in the range of low hard-band observed fluxes of $10^{-14.5} -
10^{-14.2}$~ergs~cm$^{-2}$~s$^{-1}$ and another half in the
brighter range of $>10^{-14}$~ergs~cm$^{-2}$~s$^{-1}$. This
implies that most of the Group 2 and Group 3 sources in the
field~I are expected to be extragalactic. The assumption of the
existence of some number of pre-main sequence stars without
optical counterparts\footnote{For instance, according to
theoretical PMS models CG~12 members with $0.2-0.5$~M$_\odot$ and
ages $\ga 4$~Myr will have $R > 19$~mag (= no optical
counterpart), if subject to absorption $A_V > 2$~mag.} lightly or
moderately embedded in the molecular cloud or located behind it
introduces some confusion between the simulated number of
extragalactic contaminants and the expected low-mass T-Tauri
members of the region from Group 2.

\subsection{Counterpart summary \label{cpartsum_section}}

Based on the spatial distribution of $Chandra$ sources (Figure
\ref{fig_combined_rgb}), the link between optical counterparts and
low obscuration (Figure \ref{fig_cr_vs_mede}), and the results of
these simulations of Galactic and extragalactic contaminants in
the $Chandra$ fields, we thus emerge with the following results.
With the exception of a few foreground stellar contaminants, most
of the 42 Group~1 sources ($MedE \la 1.4$~keV) are probably
lightly-obscured low-mass T-Tauri stars emerged from the CG~12
molecular cloud.  With the exception of a few hard X-ray sources
spatially coincident with the cores of the molecular cloud (\S
\ref{embedded_ages_section}), most of the 32 Group~3 sources
($MedE > 2.5$~keV) are likely extragalactic background
contaminants unrelated to the CG~12.  Among the remaining 54
Group~2 sources ($1.4 < MedE < 2.5$~keV), some can be PMS stars
embedded in the cloud or located behind it, and some can be
extragalactic background contaminants.  The 15 of these 54 with
ONIR counterparts are very likely to be YSOs.  In total, about 60
young stars are X-ray detected, increasing the known population of
the cloud by a factor of $\sim 10$.

\section{ONIR properties of $Chandra$ sources \label{ONIR_photometry_section}}

\subsection{Color-magnitude and color-color diagrams \label{onir_photometry}}

\begin{figure}
\centering
\includegraphics[angle=0.,width=7.0in]{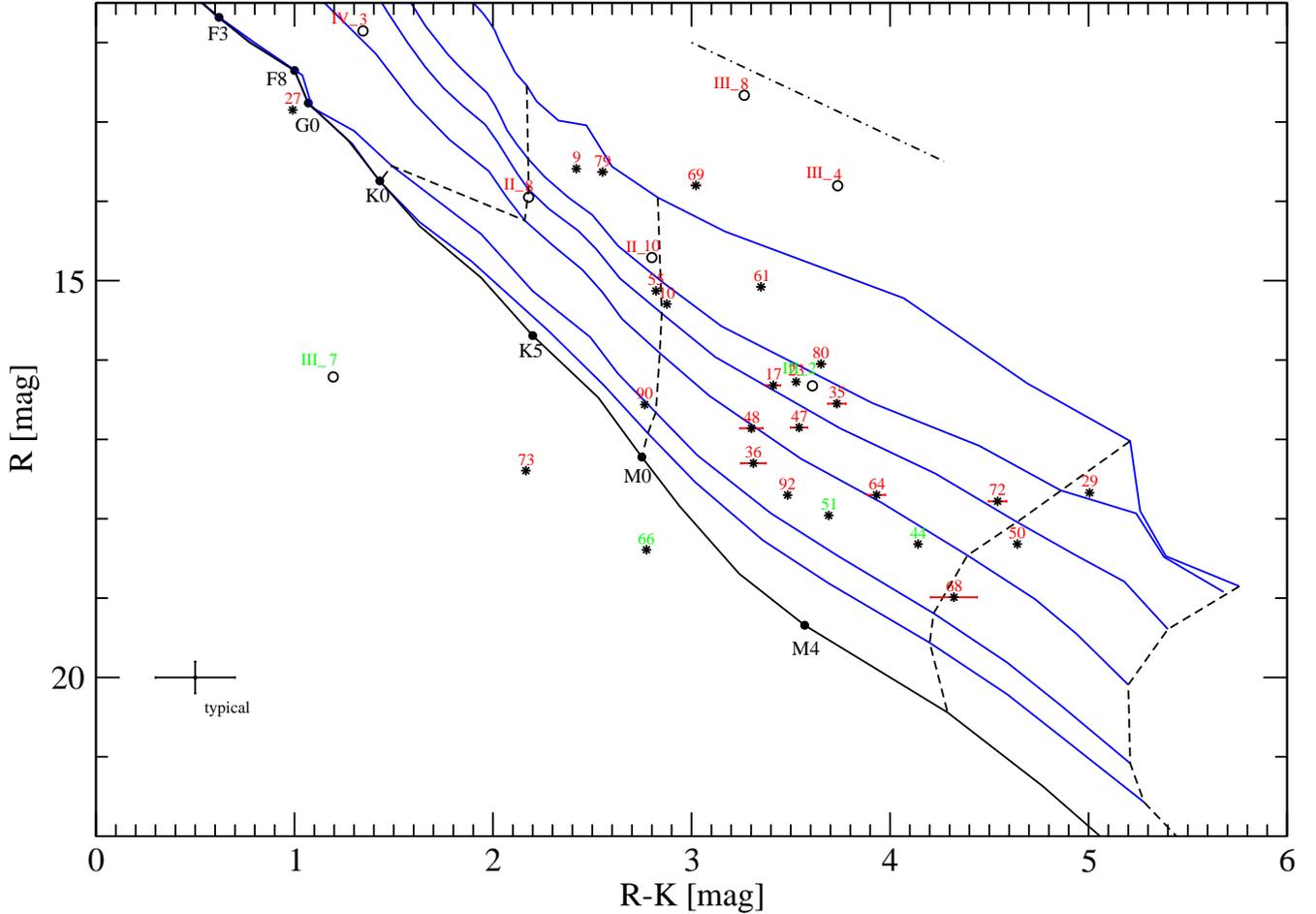}
\caption{$R$ and $K$ band color-magnitude diagram for 32 $Chandra$
sources with relatively good photometry. Black stars show sources
from Field I and open circles show sources from Fields II$-$IV.
Sources labelled in red are unabsorbed with $MedE \la 1.2$~keV
($A_V \sim 0$~mag), and sources labelled in green are lightly
absorbed with $1.3 \la MedE < 1.8$~keV ($A_V \sim 0.5-5$~mag). The
dashed-dotted line is a reddening vector of $A_V = 2$~mag. The
blue solid and dashed black lines, respectively, are PMS
evolutionary isochrones for ages 1, 3, 5, 10, 30, 50~Myr, and
evolutionary mass tracks for masses 0.1, 0.2, 0.5, and
1.0~M$_{\odot}$ \citep{Siess00}. The black solid line gives the
zero age main sequence locus.  A distance of $550$~pc to CG~12 is
assumed. \label{fig_cmd}}
\end{figure}

\begin{figure}
\centering
\includegraphics[angle=0.,width=6.0in]{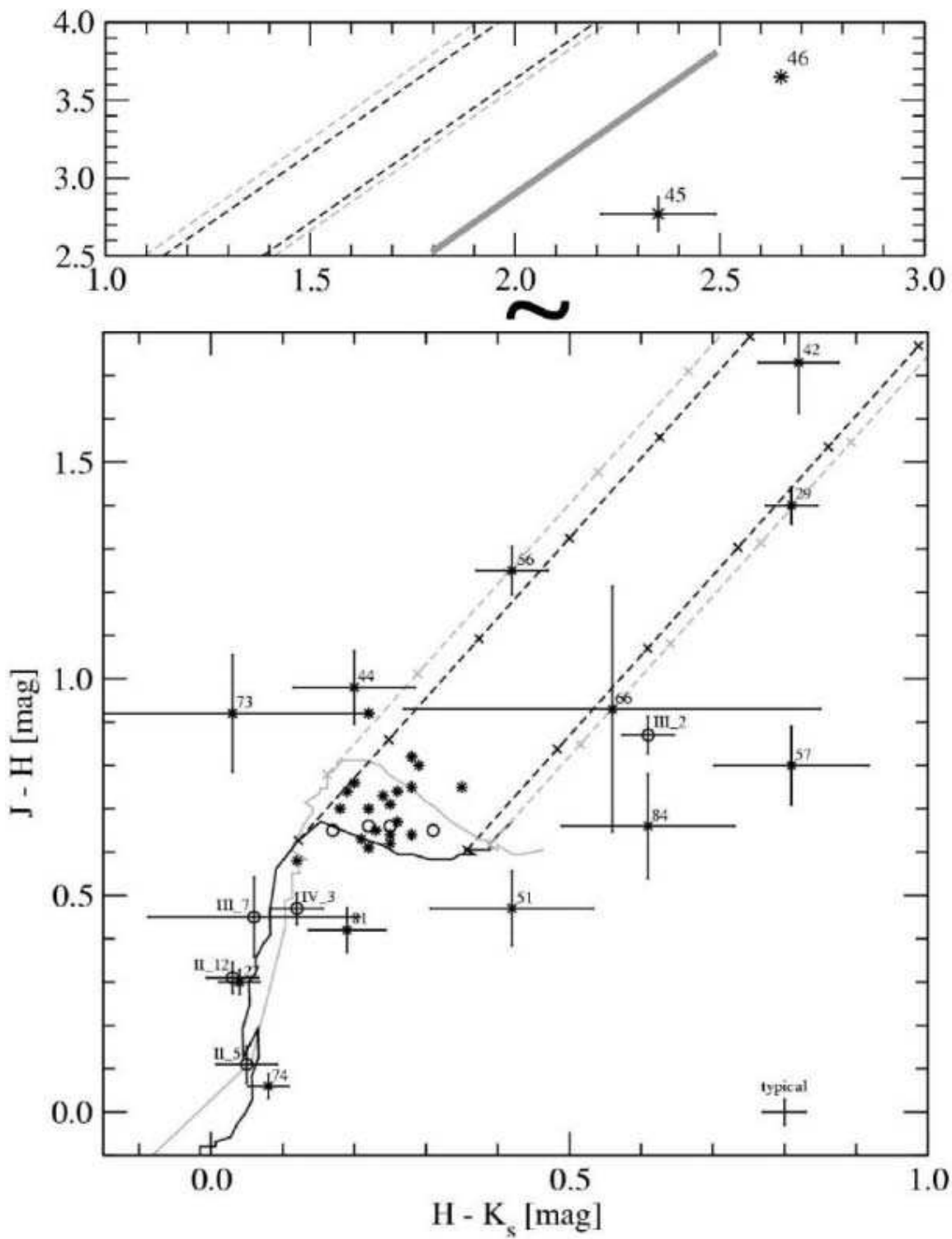} \caption{\footnotesize
$JHK_s$ color-color diagram for 43 $Chandra$ sources with reliable
NIR photometry. Primary field sources are marked as black stars
and secondary field sources as open circles.  Sources of special
interest or with large errors are labelled. Grey and black solid
lines show PMS isochrones transformed to the 2MASS photometric
system, for ages 1 and 50~Myr, respectively. Dashed lines are
reddening vectors originating from those isochrones at $\sim0.075$
and $0.75$~M$_{\odot}$, marked at intervals of $A_V = 2$~mag. The
thick grey line in the upper panel is the rightmost reddening
vector for objects from the CTT locus \citep{Meyer97}.
\label{fig_cc}}
\end{figure}

Figures \ref{fig_cmd} and \ref{fig_cc} present ONIR properties for
most of the $Chandra$ sources with ONIR counterparts based on the
photometric data in Table \ref{tbl_cntrpts}.  The $R$ $vs.$ $R-K$
color magnitude diagram in Figure \ref{fig_cmd} has 32 $Chandra$
sources with $R$-band magnitudes and high-quality $2MASS$
photometry\footnote{We restrict the sample here mostly to $2MASS$
sources with AAA photometric flags and 000 confusion flags which
have typical $JHK_s$ color errors $\la 0.03$~mag. The quality is
lower for four sources: I$_{-}$57, I$_{-}$66, I$_{-}$73, and
III$_{-}$7). $JHK_s$ magnitudes in Figure \ref{fig_cc} are in the
$2MASS$ photometric system, whereas in Figure \ref{fig_cmd} $K_s$
magnitudes have been transformed to the color system of
\citet{Bessell88} to be in agreement with $R$-band magnitudes
transformed to the \citet{Johnson66} system here.
\label{onir_footnote}}. Two lightly-obscured and two embedded
protostellar candidates are added in the $J-H$ $vs.$ $H-K_s$
color-color diagram in Figure \ref{fig_cc}: I$_{-}$29, I$_{-}$44,
I$_{-}$45 and I$_{-}$46. Several brighter sources with $R <
12$~mag, for which the photographic-CCD photometric calibration is
not known (Figure \ref{fig_mah_vs_usno}), are also added in Figure
\ref{fig_cc}: I$_{-}$57, I$_{-}$70, I$_{-}$74 and I$_{-}$81,
I$_{-}$84, II$_{-}$5 and II$_{-}$12. PMS evolutionary tracks from
the calculations of \citet{Siess00} are shown in both diagrams; a
distance of 550~pc (\S \ref{distance_section}) is assumed for the
conversion of absolute to apparent $R$ magnitude in Figure
\ref{fig_cmd}.

While accurate classification of the $Chandra$ sources must await
our upcoming optical spectroscopic measurements, the nature of the
sources with known ONIR photometry can be estimated from these
diagrams. Most are unreddened or lightly obscured ($A_V < 2$~mag),
with positions in the color-magnitude diagram consistent with
low-mass $0.2-0.7$~M$_{\odot}$ PMS stars. A wide spread of ages,
ranging from $\la 1$ to $\sim 20$~Myr, is seen. Although we have
not applied any reddening corrections, we believe this result is
robust since the reddening vector in the color-magnitude diagram
is nearly parallel to the isochrones.  No spatial pattern in ages
is seen; in particular, we find no indication that stars in the
secondary fields are systematically older than the lightly
obscured stars in the primary field.

Four stars have $JHK$ colors indicating heavy absorption
corresponding to $5 < A_V < 50$~mag: I$_-$29, I$_-$42, I$_-$45,
and I$_-$46. The latter two sources and the source I$_-$54 (with
no ONIR counterparts) are located towards the molecular cores and
are likely members of the younger embedded in the cloud population
of CG~12 (\S \ref{embedded_ages_section}).


One $Chandra$ star is unusual in its absorption characteristics.
The low $MedE = 1.2$~keV of the $Chandra$ source I$_{-}$29
indicates a low absorbing column density that stands in contrast
to the higher $A_V \sim 6$~mag inferred from its $JHK_s$ colors.
This may be due to binarity, or to an unusual disk geometry or
gas-to-dust ratio. A similar inconsistency between near-infrared
colors and X-ray absorption was found for SVS-16 in the NGC-1333
star forming cloud \citep{Getman02}.


\subsection{Inner disk emission \label{disk_emission}}

Out of 33 likely low-mass lightly and moderately obscured pre-main
sequence stars (not including members of the embedded population,
such as I$_-$45, I$_-$46, and high-mass candidate I$_-$57) in the
Figure \ref{fig_cc} color-color diagram, $3-5$ ($9-15$\%) appear
to have $K$-band excesses: I$_-$29, I$_-$51, I$_-$84, III$_-$2 and
possibly I$_-$66. Three of these stars lie on the edges of the
molecular cores, and two are far from the cloud (Figure
\ref{fig_combined_rgb}). However, the offsets from the locus of
reddened photospheres is typically small ($\sim 0.1$~mag), and the
$K$-band offsets may not be statistically significant in all
cases. In the other direction, several highly obscured sources
(established from their high values of $MedE$) are not seen by
$2MASS$, and members of the embedded population are more likely to
have heavy disks, including sources I$_-$45 and I$_-$46. We thus
view the observed $K$-band excess fraction of $\sim 9-15$\% of the
unobscured population to be only an approximation of the true
value for the full CG~12 population.

Attributing $K$-band excess to an inner dusty disk, we find that
the inner disk fraction of the CG~12 sample is intermediate
between the youngest and older known sparse young stellar
populations. For instance, the $XMM-Newton$ Extended Survey of
Taurus clouds included 65 classical and 50 weak-lined T-Tauri
stars, of which 55 and 49, respectively, were detected in X-rays
\citep{Guedel06, Telleschi06}. The Taurus cloud X-ray sample thus
has an inner disk fraction around 50\%, with most stars in the age
range $0-2$~Myr. In contrast, the $8-9$~Myr old $\eta$~Cha
cluster, which is kinematically associated with the Sco-Cen
Association and no longer has any molecular gas, has $\la 3$ out
of 15 ($\la 20$\%) low-mass members with $K$-band excesses
\citep{Lyo03, Luhman04}, although a much higher fraction is seen
at mid-infrared wavelengths \citep{Bouwman06}.  At ages around
$17-23$~Myr in the rich LCC and UCL subgroups of Sco-Cen, the disk
fraction among X-ray stars is only $\sim 1$\%, although these
disks may have been subject to OB star photoevaporation
\citep{Mamajek02}.

The $\sim 9-15$\% inner disk fraction for CG~12 stars is in accord
with our estimation that the sample has a wide range of stellar
ages with a median value around 4~Myr (\S
\ref{ages_unobscured_section}). The disk fraction is lower than
that in Taurus and higher than that in LCC/UCL Sco-Cen, and
roughly similar to that in $\eta$ Cha.

\subsection{Kinematics \label{kinematics_section}}

\begin{figure}
\centering
\includegraphics[angle=0.,width=6.5in]{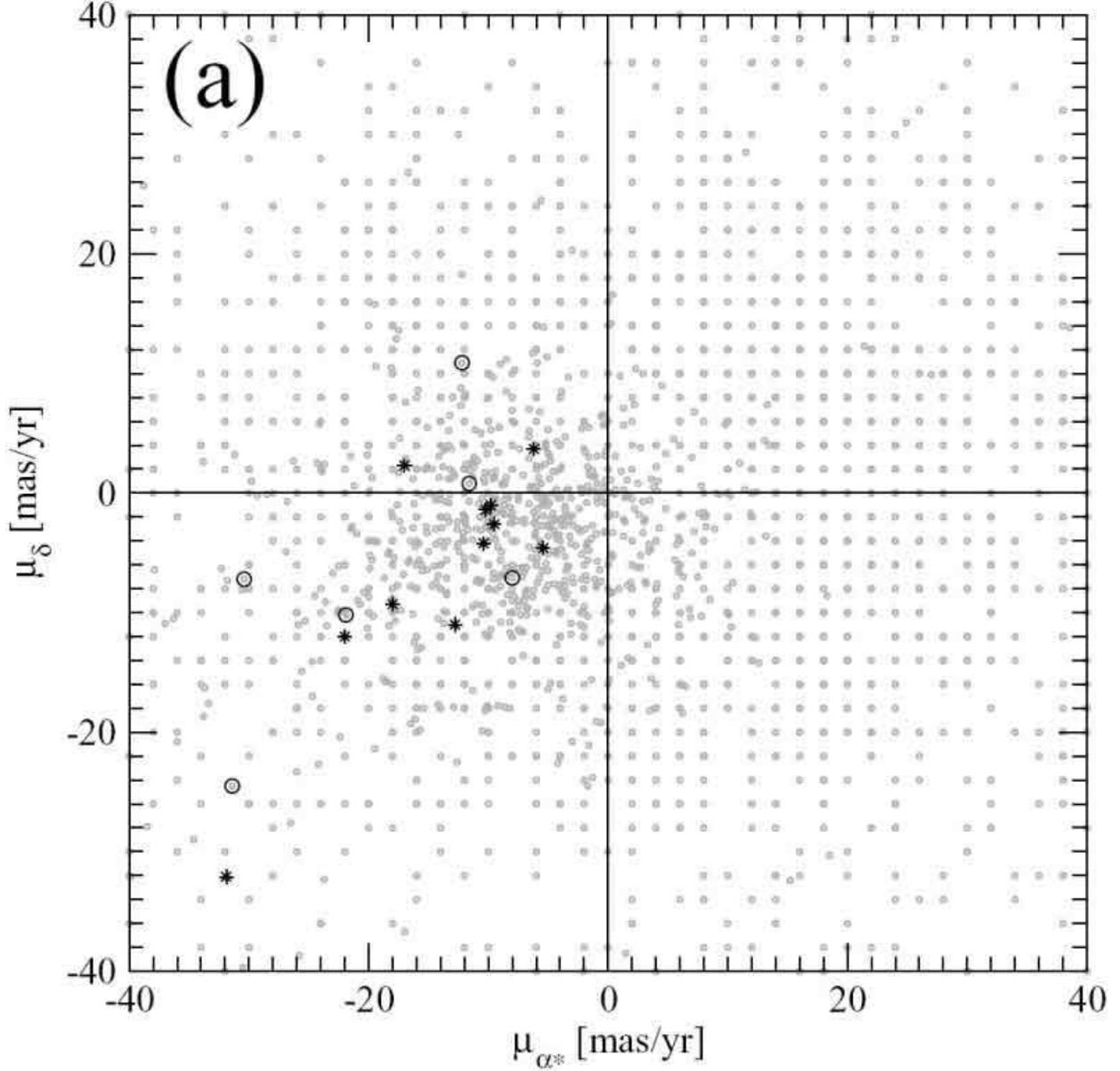}
\caption{(a) Known proper motions from the NOMAD catalog for
$>3000$ control stars inside the $Chandra$ fields (grey circles)
and for 17 $Chandra$ sources with proper motions known within $\pm
10$ mas~yr$^{-1}$ precision (11 from the primary field shown as
black $\ast$, and 6 from the secondary fields marked by black open
$\circ$). (b) Proper motions in Decl. for 23 $Chandra$ sources
($\times$) with the eight most reliable values outlined by
$\square$ and labelled, and for sources from Besan\c{c}on model
simulated within the distance of 1~kpc from the Sun (grey
$\circ$). Distances for $Chandra$ sources are randomly assigned
between $0.5-0.6$~kpc. \label{fig_prop_motions}}
\end{figure}

Proper motion data are available for 23 CG~12 X-ray sources (Table
\ref{tbl_cntrpts}) and can, in principle, be important for
confirming membership in the cloud population and studying its
dynamical evolution. CG~12 would have an unusual space motion, for
example, if it were related to Galactic high-velocity clouds
\citep[e.g.][]{Odenwald92}. Best known proper motions from
catalogs compiled in NOMAD are given in Table \ref{tbl_cntrpts}.
Twenty three X-ray sources have known proper motions, but some
values are quite uncertain. The vector point diagram in Figure
\ref{fig_prop_motions}a plots the 17 $Chandra$ stars with proper
motions known within $\pm 10$ mas~yr$^{-1}$ precision.  They are
compared with $>3000$ control sources taken from the NOMAD catalog
within the $Chandra$ fields. Most of the optically bright
$Chandra$ stars have proper motions in the field star
concentration around $(\mu_\alpha*,\mu_\delta) \simeq (-10,-5)$
mas~yr$^{-1}$.  However, a few appear to be outliers, such as
I$_-$57 and III$_-$8 around (-30,-30).

In Figure \ref{fig_prop_motions}b, we compare one component of the
observed proper motions to a simulated sample of nearby ($d \leq
1$~kpc) field stars using the Besan\c{c}on stellar population
model (see \S \ref{contamination_section}). This shows that the
distribution of $Chandra$ stellar proper motions is typical of
those of field stars in the direction of CG~12 located at
distances of $0.4 \la d \la 1$~kpc. Thus, the proper motions of
X-ray selected young stars in CG~12 can be attributed entirely to
the established motions in the Galactic disk with no evidence for
peculiar space motions.

\section{Distance to CG~12 \label{distance_section}}

The distance to CG~12 is controversial, with estimates ranging
from 100 to 600~pc \citep{Maheswar04}. The closer estimates arise
from assumptions of a small scale-height for the local
interstellar medium \citep{vanTill75,Keto86}. \citet{Bourke95b}
assigned a distance of 400~pc assuming an association with the
Vela-Gum region. From the optical photometry of intermediate-mass
stars in the cloud, \citet{Williams77} and \citet{Marraco78}
obtained distance estimates around 600~pc which are commonly used
by other researchers. \citet{Maheswar04} obtained ONIR photometry
of a hundred stars around the globule to derive a distance to
CG~12 of 550~pc assuming they are main sequence field stars. Here
we consider the distance range $300 < d < 630$~pc and use X-ray
properties of the PMS members to constrain the distance further.

The proposed here method of distance determination is based on two
interlocking features: PMS X-ray luminosities depend on stellar
mass; and X-ray luminosities and masses scale differently with
assumed distance.  While it is well-established that most PMS
X-rays seen in the $Chandra$ band arise from violent magnetic
reconnection flares above the stellar surface \citep{Feigelson07},
it is not fully understood why it scales with stellar mass.  It is
quite possible that the observed $L_x-M$ correlation arises
indirectly from two physical effects: a link between mass and
stellar radius, and a link between X-ray emission and stellar
volume or surface area \citep{Preibisch05b,Telleschi06}.  In any
case, the $L_x-M$ correlation has been seen in a wide variety of
PMS populations independent of the detailed age distribution,
star-formation mode, cluster richness.

Some care is needed in the definition of the $L_x-M$ relation:
estimates of the X-ray luminosities and masses,  and the treatment
of scatter in the $L_x-M$ diagram differ in various studies. In
our past $Chandra$ studies, we defined the $L_x-M$ relationship
using the quantity $L_{t,c}$, the absorption-corrected luminosity
in the ``total" $0.5-8$~keV band (see Getman et al. 2005 and notes
in Table \ref{tbl_thermal_spectroscopy} below). The lightly
obscured Orion COUP sample gives the regression lines $L_{t,c} =
1.44 \log M + 30.37$ and $L_{t,c} = 1.13 \log M + 30.34$ erg
s$^{-1}$ for masses estimated from optical spectroscopy and
theoretical evolutionary tracks of \citet{Siess00} and
\citet{Palla99}, respectively \citep{Preibisch05b}.
\citet{Getman06} obtained $L_{t,c} = 1.99 \log M + 30.43$ erg
s$^{-1}$ for the Cepheus OB3b and Orion stars using NIR photometry
and Siess et al. tracks. The $XMM-Newton$ XEST survey, using X-ray
luminosity defined in the broader $0.3-10$~keV band, and a
bisector line rather than an ordinary least-squares regression
line, gives $L_{XMM} = 1.91 \log M + 30.44$ erg s$^{-1}$ using
combined stellar spectroscopic results of many studies and Siess
et al. tracks \citep{Telleschi06}. We adopt here the
Orion-Cepheus~OB3b regression line from \citet{Getman06}, which
uses stellar mass estimate methods more comparable to that of the
current work, although recognizing that all the regression lines
cited above lie very close to each other.

To estimate the distance to CG~12, we restrict our consideration
to a highly reliable subsample of X-ray bright, completely
unobscured ($MedE < 1.2$~keV and $\log N_H \la 20$~cm$^{-2}$)
X-ray sources with $L_{t,c}$ values given in Table
\ref{tbl_thermal_spectroscopy} and accurate photometry in Table
\ref{tbl_cntrpts}. Masses are estimated from the $R$ $vs.$ $R-K$
color-magnitude diagram (Figure \ref{fig_cmd}) using
\citet{Siess00} tracks.  The sample includes twelve likely T-Tauri
stars without K-band excesses of which ten are from the primary
$Chandra$ field (I$_{-}$9, 10, 17, 23, 55, 61, 64, 69, 79, and
90), and two are from the secondary fields (II$_{-}12$ and
IV$_{-}$3)\footnote{Two more X-ray unobscured stars from the
secondary fields, III$_{-}$4 and III$_{-}$8, are not included here
due to problems with their estimated ages (see \S
\ref{ages_unobscured_section}).}. Errors on the estimated
photometric masses for the stars in the sample are derived from
Monte-Carlo simulations based on the observed photometric
uncertainties convolved with the Siess et al. evolutionary tracks.
Errors on $L_{t,c}$ are estimated through Monte-Carlo simulations
described in \S 4.3 of \citet{Getman06}.

\begin{figure}
\centering
\includegraphics[angle=0.,width=6.5in]{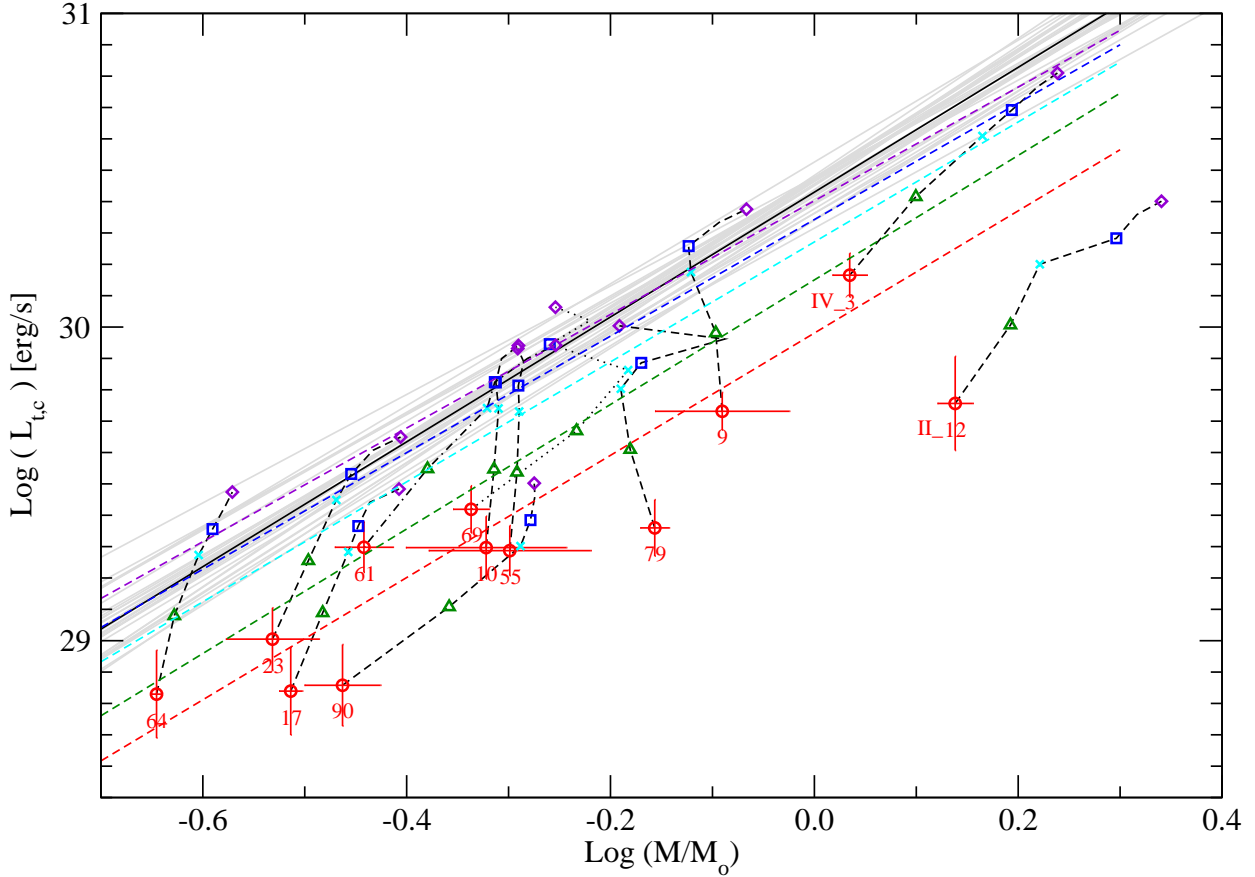}
\caption{X-ray luminosity ($L_{t,c}$) plotted against stellar mass
for 12 unobscured X-ray T-Tauri stars, assuming 5 different
distances to CG~12: $\bigcirc$ and red line for $d=300$~pc;
$\triangle$ and green line for $d=400$~pc; $\times$ and cyan line
for $d=500$~pc, $\square$ and blue line for $d=550$~pc; $\diamond$
and violet line for $d=630$~pc. Regression lines exclude source
\#12. Red error bars indicate uncertainties on $\log L_{t,c}$ and
$\log M/M_{\odot}$ at $d=300$~pc, and are similar for other
distances. The solid black line is a regression fit for PMS stars
in the Orion Nebula Cluster with standard deviation shown by the
grey lines. \label{fig_lx_vs_mass}}
\end{figure}

Figure \ref{fig_lx_vs_mass} shows the resulting X-ray luminosities
and photometric masses of these stars calcualted for a range of
possible distances to the cloud. It is clear from the figure that
the assumed distances $\la 500$~pc (cyan, green, red symbols) are
insufficient to place CG~12 PMS objects on the Orion $L_x-M$
regression line (black and grey solid lines). Distances of
$550-630$~pc (blue and violet symbols) are clearly preferable. The
lower distance of $550$~pc is further preferred if the recent
revision to the Orion Nebula distance by \citet{Jeffries07} is
correct\footnote{An analysis based on the rotational properties of
low-mass PMS ONC stars with exclusion of likely accreting systems
suggests a distance of $390 \pm 30$~pc, substantially closer than
the commonly used $440 \pm 30$~pc.}. Throughout our study, we
adopt the distance to CG~12 of $550$~pc.

We thus independently confirm that the distance to CG~12 is $>
500$~pc and more likely is close to $550$~pc, in close agreement
with the measurement by \citet{Maheswar04} based on photometry and
reddening of foreground and background field stars.  An important
consequence is the confirmation that the cloud is indeed located
at a height of $\ga 200$~pc above the Galactic plane. The
existence of any star forming dark cloud at such Galactic location
is unusual, and the cometary structure of CG~12 in such an
isolated environment is even more mysterious (\S
\ref{discussion_section}).

\section{The CG~12 stellar population \label{stellar_population_section}}

\subsection{Stellar census \label{stellar_census_section}}

Five intermediate-mass stars (three B and two A stars), and
several FG stars had been identified as likely members of the
CG~12 region in previous ONIR studies \citep{Williams77}.  The
three B stars in the region are detected by $Chandra$.  The $V
\sim 11$~mag B4 and $V \sim 10$~mag B7 components of the double
star Herschel~4636, which illuminates the reflection nebula
NGC~5367 \citep[= GN~13.54.7][]{Williams77}, correspond to
$Chandra$ sources I$_{-}58$ ($\log L_{t,c} \simeq 29.7$ erg
s$^{-1}$) and I$_{-}57$ ($\log L_{t,c} \simeq 30.1$ erg s$^{-1}$),
respectively. The $V \sim 10$~mag B8 star CD~-39$^{\circ}$8583
associated with the reflection nebula GN~13.54.9 is I$_{-}74$
($\log L_{t,c} \simeq 30.2$ erg s$^{-1}$ erg s$^{-1}$). The two A
stars\footnote{These are sources \#\# 6 and 8 from
\citet{Williams77}, typed as A4 and A2 by \citet{Maheswar04}.
Source \#8 may be powering the reflection nebula B77~146
\citep{Bernes77}, located just northwest of the cloud (Figure
\ref{fig_combined_rgb}$b$).} have emission fainter than $\log
L_{t,c} \la 29.0$ erg s$^{-1}$, the sensitivity level in our
primary $Chandra$ field.

These X-ray properties of the CG~12 intermediate-mass stars are
consistent with past studies on A and B star samples. These stars
lack both convection zones which drive surface magnetic flares and
the strong radiation-driven stellar winds which drive
high-temperature shocks. It is suspected that their intrinsic
X-ray emission is low or absent, and some appear in $Chandra$
images only when accompanied by a lower mass T-Tauri companion
\citep[e.g.][]{Stelzer06}.

The important new $Chandra$ result is the identification of nearly
60 low-mass young stars, increasing the known population of the
cloud by nearly a factor of 10 (\S \ref{counterparts_section} and
\ref{contamination_section}). Three of these are very likely
embedded protostars (\S \ref{embedded_ages_section}), while the
rest are lightly-moderately obscured T-Tauri stars. As outlined in
\S \ref{introduction_section}, the existence of these stars is
expected from an extrapolation of a standard Initial Mass Function
from the previously known intermediate-mass members.  The five
known BA stars demand the presence of $\sim 80$ T-Tauri stars
within the mass range of $0.2-2$~M$_{\odot}$ assuming the Galactic
field IMF of \citet{Kroupa02}.

With the sensitivity limits of our X-ray observations around $\log
L_{t,c} \simeq 29.0 (30.0)$ erg s$^{-1}$ in the primary
(secondary) fields, we expect our samples to be sensitive down to
$\sim 0.2$ (0.5) M$_\odot$, and complete at masses higher than
these. This inference emerges from the established statistical
correlation between X-ray luminosity and stellar mass
\citep{Preibisch05b, Telleschi06}. Thus, our X-ray detected sample
constitutes most, but not all, of the expected low mass
population.

We can estimate the star formation efficiency (SFE) of CG~12,
defined as the ratio of integrated stellar mass to the total mass
of molecular gas and stars. The 80 stars in the range $0.2<M<2$
M$_\odot$ have a total mass around $30-35$ M$_\odot$, and the
intermediate-mass stars have a total mass around 20 M$_\odot$,
given a stellar mass around 50 M$_\odot$. The mass of the observed
molecular gas is estimated to be $100-250$~M$_{\odot}$ (\S
\ref{introduction_section}). The implied SFE is then in the range
15\%-35\%. This should be considered an upper bound on the SFE
since the original cloud was probably more massive prior to
ablation into a cometary morphology. These values are typical
among other star forming regions \citep[][and references
therein]{Elmegreen00}. However, it is significantly higher than
the SFE we found for the much smaller and younger cometary globule
IC~1396N (typical among hundred of known BRCs) where the SFE is
estimated to be less than $4$\% based on the $Chandra$ sample
\citep{Getman07}. This suggests that the triggering mechanisms in
CG~12 may differ from smaller globules on the edges of HII
regions.

\subsection{Ages of the Unobscured Population \label{ages_unobscured_section}}

Stellar ages for the unobscured low-mass ($\la 2$~M$_{\odot}$)
X-ray population of CG~12 can be estimated from the
color-magnitude diagram in Figure \ref{fig_cmd}. These stars have
$MedE \la 1.2$~keV and $R \ga 12$~mag located above the ZAMS.
Three stars are excluded from consideration and require more
study: I$_{-}$29 due to inconsistency between its X-ray and NIR
obscuration (\S \ref{ONIR_photometry_section}), and III$_{-}$4 and
III$_{-}$8 because their inferred ages are characteristic of
Class~0/I protostars rather than visible T-Tauri stars (although
they have normal colors in the $JHK_s$ color-color diagram, Figure
\ref{fig_cc}). Thus, 21 stars are considered in this analysis:
I$_-$9, 10, 17, 23, 35, 36, 47, 48, 50, 55, 61, 64, 68, 69, 72,
79, 80, 90, 92 from the primary field, and II$_{-}$8 and
II$_{-}$10 from the secondary fields.

\begin{figure}
\centering
\includegraphics[angle=0.,width=6.5in]{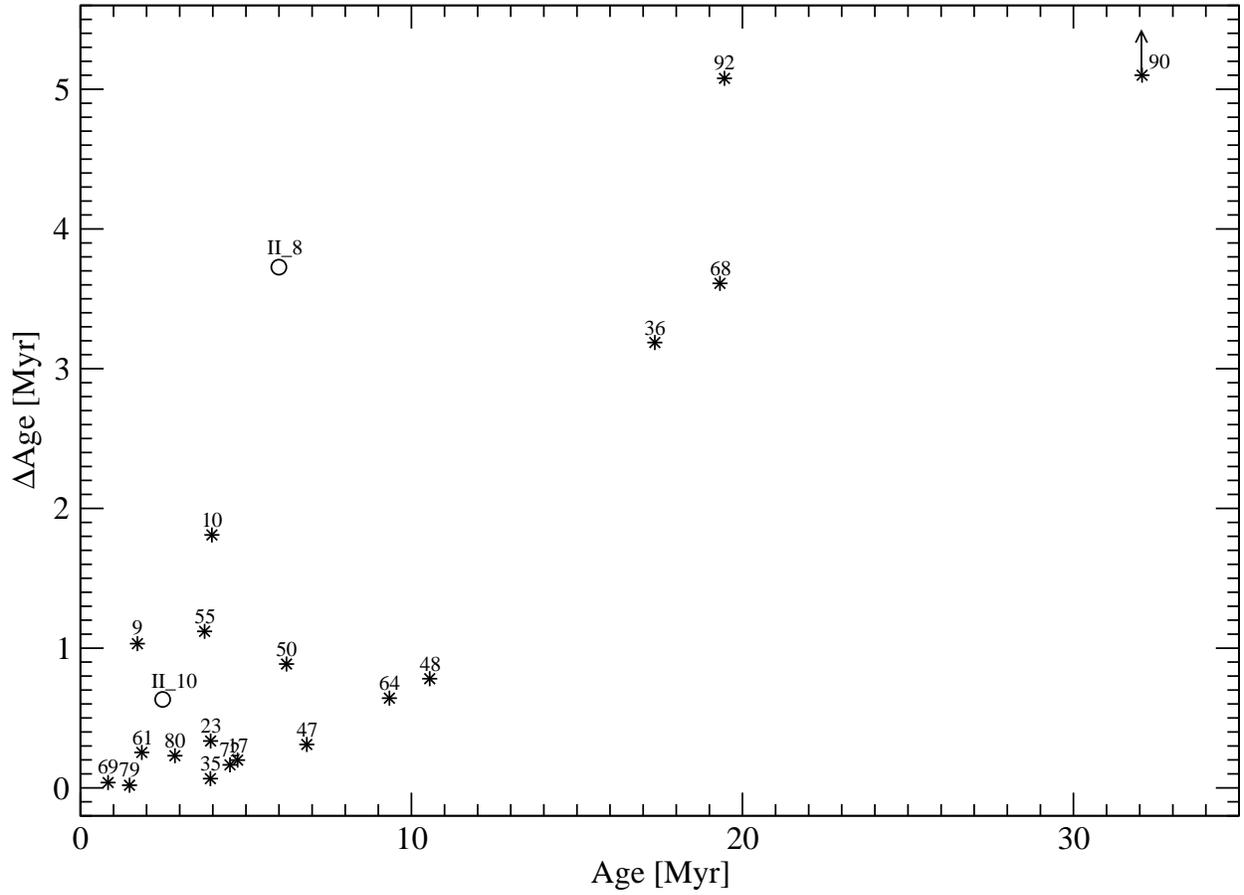}
\caption{Photometric isochronal age estimates and their formal
errors for the unobscured population of low-mass X-ray stars in
CG~12 for primary field ($\large \star$) and secondary field
($\large \circ$) sources. \label{fig_dage_vs_age}}
\end{figure}

The derived photometric ages for these stars using \citet{Siess00}
PMS models in $R$ vs. $R-K$ (which are similar to $R$ vs. $R-H$)
are shown in Figure \ref{fig_dage_vs_age} along with their formal
errors, $\Delta$Age. The errors were derived similarly to the mass
errors in \S \ref{distance_section} using Monte Carlo simulations
of photometric magnitudes and colors according to each source's
reported photometric errors, and placing simulated stars on PMS
isochrones. The errors do not account for a possible presence of
unresolved binaries or variability
\citep{Preibisch99,Burningham05}. Unresolved binaries would tend
to make objects appear up to a factor of 2 brighter, leading to an
underestimation of the true age. PMS stars are known to be
variable due to rotational modulation of starspots and, for
classical T-Tauri stars, due to variations in accretion.  Typical
starspot fluctuations are $\Delta$R$\simeq \Delta$K$\simeq
0.1$~mag \citep{Carpenter01,Herbst02} and accretion variations can
exceed 1~mag \citep{Stassun06}. None of the 21 low-mass stars used
in this analysis possess noticeable $K_s$ excesses (Figure
\ref{fig_cc}), so fluctuations likely do not exceed $0.1$~mag. But
variability often moves the star parallel to isochrones and may
not contribute significantly to apparent age spreads
\citep{Burningham05}. If we add this additional uncertainty to
photometric magnitudes and colors of CG~12 sources, the inferred
age uncertainties increase by factors of $\sim 2-3$. We denote the
derived ages as ``photometric isochronal ages'' in order to
distinguish them from the true stellar ages.  Our upcoming
spectroscopic study should provide more reliable estimates of
individual CG~12 stellar ages.

The distribution of the photometric isochronal ages in Figure
\ref{fig_dage_vs_age}, combined with the presence of several
embedded protostars not plotted here (\S
\ref{embedded_ages_section}), indicates a large age spread from
$<1$ to $\sim 20$~Myr. The age spread appears real and cannot be
attributed to photometric errors.  There may be two or more
distinct populations in CG~12: a very young embedded population
forming in the molecular cores right now (\S
\ref{embedded_ages_section}) as a part or independent of a younger
and richer population of T-Tauri stars with ages $1-10$~Myr,
median value around $\sim 4$~Myr (mean around $4.5$~Myr) and a
standard deviation of 2.7~Myr; and a sparse older population of
likely $15-30$~Myr stars. The latter sub-sample (I$_{-}$: 36, 68,
92, and 90) also has $JHK_s$ photometric colors close to the
50~Myr (older) isochrone in Figure \ref{fig_cc}. However, the
results are also consistent with a continuous star formation
history.

We note that age spreads from 0 to 30~Myr have been reported in a
number of young stellar populations. Most relevant to this study,
\citet{Ogura06} find mean ages of $1-2.5$~Myr and standard
deviations of $0.6-2.5$~Myr, with two stars around 5~Myr, in the
stellar populations around four CGs. Their sample is derived from
ONIR techniques that are less effective than $Chandra$ images in
locating older weak-lined T-Tauri stars, so the true maximum age
may be older than they estimate.


Broad stellar age distributions are found in other well-studied
star forming regions.  \citet{Kenyon95} find a smooth age
distribution from $<1$ to $>10$ Myr in the Taurus-Auriga clouds,
although \citet{Hartmann03} \citep[see also][]{Slesnick06} argues
that the older stars are unrelated to the clouds.  About 10\% of
the Orion Nebula Cluster membership have ages around 10~Myr with
the remainder much younger around $\la 1$~Myr
\citep{Slesnick04,Palla05,Palla07}. The young stellar cluster
associated with $\sigma$~Ori has an age spread up to $\sim 30$~Myr
with median age of $\sim 4$~Myr
\citep{Oliviera02,Oliviera04,Sacco06}. The Chamaeleon I population
shows an age range from $<1$ to 10~Myr \citep{Luhman04}. In
contrast, no age spread is found in the $\sim 5$~Myr old Upper-Sco
OB association \citep{Preibisch99}. From a theoretical
perspective, researchers are actively debating whether molecular
clouds are short-lived and produce brief bursts of star formation
\citep[e.g.][]{Elmegreen00,Hartmann01,Ballesteros-Paredes07} or
whether clouds are long-lived and produce stars in a more
continuous fashion \citep[e.g.][]{Palla00, Mouschovias06,
Krumholz07}.

\subsection{Ages of the Embedded Population \label{embedded_ages_section}}

Signs of ongoing star formation in CG~12 include a bipolar
molecular outflow \citep{White93} and its millimeter-continuum
candidate source in the CG~12S molecular core \citep{Haikala06},
and the far-infrared source IRAS~13546-3941 in CG~12N, resolved in
the near-infrared as a Class~I visual binary protostellar system
with 5\arcsec\/ separation  \citep{Santos98}. Here we describe the
$Chandra$ detection of the components of the NIR visual
protostellar binary in CG~12N (I$_{-}$45 and I$_{-}$46), and
present evidence for a newly discovered protostar in CG~12S
(I$_{-}$54).

$Chandra$ sources I$_{-}$45 and I$_{-}$46 exhibit absorptions of
$\log N_H \sim 22.1-22.2$~cm$^{-2}$ (Table
\ref{tbl_thermal_spectroscopy}) which correspond to $A_V \sim
8-10$~mag assuming a standard gas-to-dust ratio from
\citet{Vuong03}.  This is consistent with a location near the
center of the molecular core CG~12N (Figure
\ref{fig_combined_rgb}$b$) which has a total absorption of $\sim
10-13$~mag \citep{Haikala06}. Their NIR colors are consistent with
Class~I objects (Figure \ref{fig_cc}), although an early Class~II
phase is possible. Their X-ray luminosities are similar, $\log
L_{t,c} \sim 30.3-30.5$ erg s$^{-1}$, and are consistent with the
brightest known solar-mass T-Tauri stars and many Class~I
protostars \citep{Imanishi01,Wolk05} and are definitely the
strongest among the hard X-ray sources seen through the CG~12
cloud.

Among all CG~12 $Chandra$ sources with $NetCts > 20$ (Table
\ref{tbl_thermal_spectroscopy}), source I$_{-}$54 has the hardest
spectrum ($MedE \sim 3.2$~keV), but it is not identified with any
known ONIR source. Seen in projection against the CG~12S molecular
peak (Figure \ref{fig_combined_rgb}$b$) where the absorption
through the cloud is $\sim 6-8$~mag \citep{Haikala06}, its X-ray
spectrum shows a much higher absorption of $\log N_H \simeq 22.6$
cm$^{-2}$ corresponding to $A_V \sim 25$~mag. The additional
absorption of $\sim 20$~mag that cannot be attributed to the cloud
likely originates in a protostellar envelope.  In IC~1396N,
\citet{Getman07} show from a $Chandra/Spitzer$ comparison that
such heavy absorption arises only in local protostellar envelopes,
and in the Serpens cloud \citet{Giardino07} find absorptions for
all Class~I protostars to be $\log N_H > 22.3$ cm$^{-2}$. I$_-$54
has an X-ray luminosity similar to the other two Class~I
protostars in CG~12, $\log L_{t,c} \simeq 30.3-30.7$ erg s$^{-1}$,
higher than most of the unobscured T-Tauri stars. A similar trend
is seen in IC~1396N and Serpens. Finally, I$_-$54 is one of only
two sources (along with the unobscured intermediate-mass star
I$_{-}$70) that show significant X-ray variability (Table
\ref{tbl_bsp}). This is likely due to a magnetic reconnection
flare during our short $Chandra$ exposure. This feature excludes
the possibility that I$_-$54 might be an extragalactic
contaminant; extragalactic sources, even active galactic nuclei,
of this brightness do not exhibit variability over an 8 hour
exposure \citep[e.g.][]{Bauer03, Paolillo04}. Source I$_-$54 thus
shows all of the X-ray characteristics of a Class~I protostar:
hard spectrum, high absorption, high luminosity, and magnetic
flaring.

We thus confirm that the CG~12 molecular globule harbors the
population of very young objects with ages of $\la 0.1$~Myr,
typical of Class~I protostars.

\subsection{Spatial structure \label{spatial_structure_section}}

The $Chandra$ results show that the young stars are much more
widely distributed than today's cloud (Figure
\ref{fig_combined_rgb}$a$). Previously noted cloud members, and
our concentrated group of obscured X-ray stars, lie within the
CG~12 cloud in a $<1 \times 1$~pc$^2$ region. But the unobscured
young X-ray emitting population is distributed widely over the
four $Chandra$ fields which subtend $5 \times 5$~pc$^2$ (Figure
\ref{fig_combined_rgb}$a$).  The concentration of X-ray sources
inside the inner portion of the $ROSAT$ detector (Figure
\ref{fig_rosat}), which subtends a much greater area than
$Chandra$'s detector, shows that most of the X-ray population lies
within the $Chandra$ fields.

\section{Discussion \label{discussion_section}}

\subsection{Implications for CG star formation models}
\label{age_discussion_section}

In \S \ref{ages_unobscured_section} and
\ref{embedded_ages_section}, we establish that the CG~12 globule
produced young stars over a period of $\ga 10$~Myr, most of which
likely formed around $4.5 \pm 2.7$~Myr ago.  Many of these stars
are now widely distributed around the cloud. The cloud also
harbors three very young embedded protostars forming today in its
molecular cores.

These findings are similar to age distributions found in other
low-mass star forming regions such as Taurus-Auriga
\citep{Kenyon95} and Chamaeleon I \citep{Luhman04} where stellar
ages range over $<1$~Myr to $\ga 10$~Myr.  These broad age ranges
are inconsistent with rapid star formation scenarios where
gravitational collapse occurs in one or two turbulent crossing
times \citep{Elmegreen00, Hartmann01, Ballesteros-Paredes07}. For
a cloud like CG~12 with size $1-2$~pc (although prior to ablation
it may have been larger), the characteristic large-scale turbulent
velocity will be $\sim 2$~km s$^{-1}$ \citep{Larson81} with
corresponding crossing time $\sim 2$~Myr. These rapid star
formation models thus predict that all cloud members should have
ages $\la 2$~Myr.  Instead, the CG~12 age distribution suggests
either a more continuous star formation process \citep{Palla00,
Mouschovias06, Krumholz07, Nakamura07}, or possibly several
episodes of a rapid star formation. The age distribution in CG~12
is consistent with an exponentially accelerating star formation
rate \citep{Palla00} providing the characteristic timescale is
several million years.

The broad age spread is also inconsistent with the rapid timescale
(few $\times 10^5$~yr) predicted by simple triggered star
formation models of CGs outlined in \S \ref{introduction_section}.
However, more complex models of shock triggered star formation
might explain the age distribution of CG~12 stars. Two
possibilities can be considered. First, RDI simulations producing
only brief durations of star formation \citep{Lefloch94,Miao06}
modelled only small globules with initial sizes $<1$~pc typical
for known bright rimmed clouds \citep{Sugitani91} with volumes
much smaller than the initial size of the atypically large CG~12
cloud. The RDI shock front may thus take millions of years to
propagate through the whole CG~12 cloud during which distributed
molecular cores could be compressed, triggering localized star
formation at different times. Second, unlike triggering by the
continuous flux of photoevaporating ultraviolet light from OB
stars in HII regions, the triggering of CG~12 could arise from a
sequence of distinct episodes separated by millions of years.
These triggers could be the shocks of distinct supernova from a
massive OB association. Astronomical evidence for these scenarios
is discussed in \S \ref{origin_section}.

The wide spatial distribution of young stars, roughly $5 \times 5$
pc$^2$ around a cloud with current size $1 \times 2$ pc$^2$, has
additional implications for the star formation process. It
suggests either that the cloud was originally much larger before
it was ablated into a cometary morphology, or that the stars
drifted from the core, or both. The members of the unobscured
population of CG~12 are uniformly distributed across the field and
do not show any spatial gradient in their derived photometric
stellar ages that could be indicative of sequential star formation
produced by a single triggering mechanism such as a propagating
shock. A spatio-temporal gradient of this type has recently been
seen in several other CGs and bright-rimmed clouds
\citep{Matsuyanagi06, Getman07, Ogura06}. Star drifting, on the
other hand, requires a stellar velocity dispersion around 1 km
s$^{-1}$ which can reasonably arise either from dynamical
interactions during star formation \citep{Bate03} or from
turbulent motions in the natal gas \citep{Feigelson96}.

The absence of an obvious age gradient in CG~12 is not yet
definitive, as our optical spectroscopy and mid-infrared $Spitzer$
observations have not been completed. The addition of a spatial
distribution of infrared-bright low mass stars with massive disks,
as found from $Spitzer$ IRAC maps of IC~1396N by \citet{Getman07},
may give a crucial new clue and could establish the validity of
the expected cloud ablation and sequential triggered star
formation mechanism.

\subsection{New observational keys to the origin of CG~12
\label{origin_section}}

With a strange location and orientation, and no reported external
hot OB star further from the plane illuminating and ablating the
globule, the origin of CG~12 is very uncertain. The best attempted
explanation was made 30 years ago by \citet{Williams77} who
suggested that a large expanding high-latitude interstellar shell
could have traversed CG~12 7-19~Myr ago.  This shell is now called
GC~323+34-021 in the Leiden-Dwingeloo H~I survey
\citep{Ehlerova05}\footnote{Due to the limiting coverage of the
Leiden-Dwingeloo H~I survey, \citet{Ehlerova05} report only the
top part of the shell.} and GIRL~G318+32 in the IRAS loop catalog
\citep{Konyves07}. Produced by one or more supernova explosions
from a still-unidentified association, the encounter would have
drawn out the 10~pc-long tail and triggered the star formation in
the globule. This explanation suffers a major difficulty: a single
supernova explosion (perhaps from a runaway O star ejected from a
Galactic Plane cluster) can not be responsible for such a large
$\sim 20^\circ$ cavity, while a superbubble produced by many
supernova remnants requires a rich high-latitude stellar cluster
which has not yet been identified.

\begin{figure}
\centering
\includegraphics[angle=0.,width=6.0in]{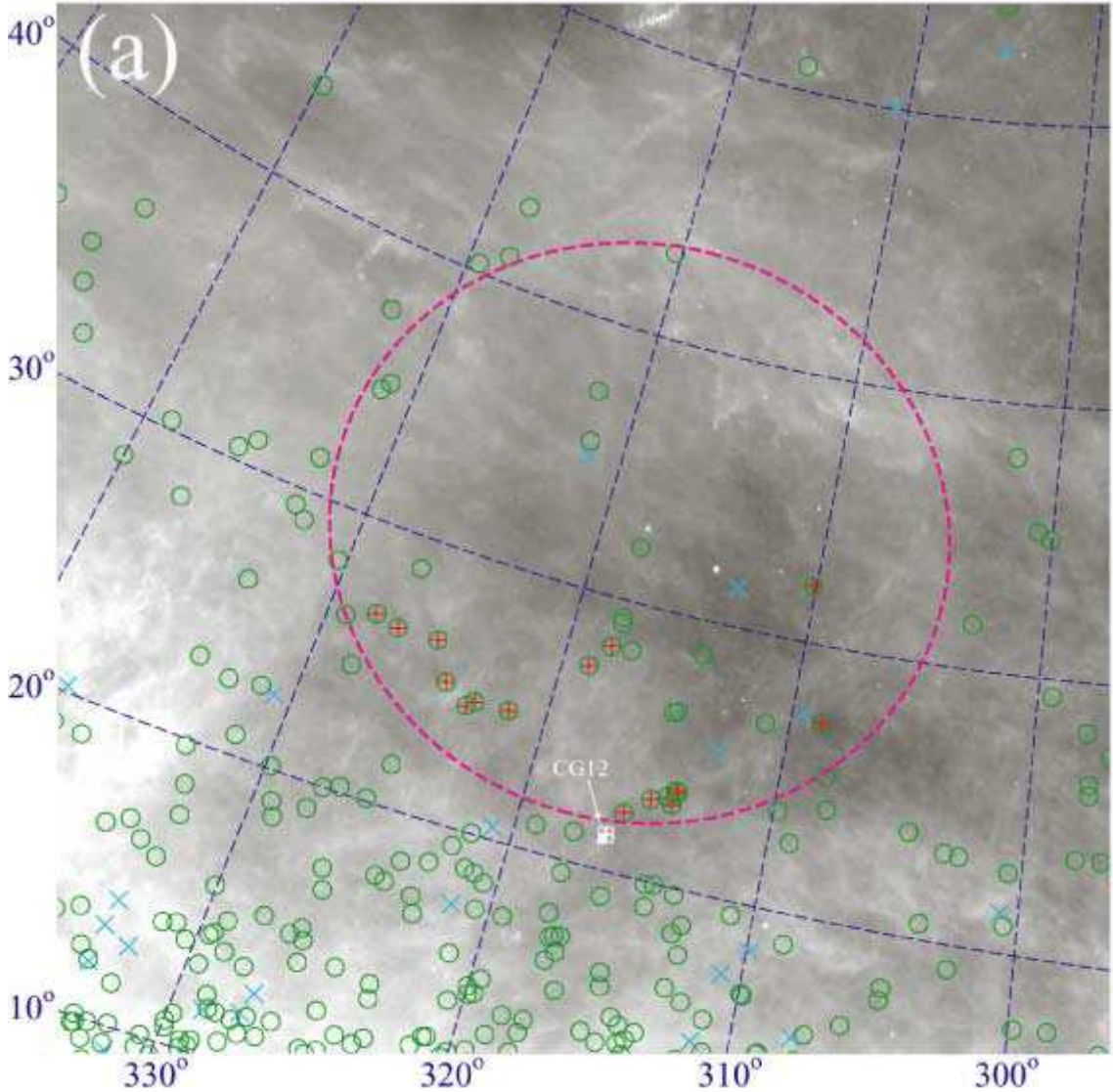}
\caption{(a) IRAS~100$\mu$m image ($\sim 40\arcdeg \times
40\arcdeg$) with the $Chandra$ fields marked as small white
squares. The wall of the giant IRAS loop GIRL~G$318+32$ is marked
by the magenta circle. B-type stars from the HD catalog are marked
by green circles, and those inside the loop with estimated
photometric distances of $350<d<700$~pc from the Sun are indicated
by red $+$ symbols. Radio pulsars are shown with cyan $\times$.
(b) Close-up view ($\sim 3.5\arcdeg \times 3.5\arcdeg$) of CG~12
and a nearby group of B stars from DSS: A -- HD~120958, B --
HD~120124, C -- HD~119627, D -- HD~119484, E -- HD~119338, and F
-- HD~119277. The tail axis of the CG~12, marked as white dashed
line, passes right through HD~120958. \label{fig_origin}}
\end{figure}

To further test this model, we searched for intermediate-mass
B-type members of the hypothetical high-latitude young stellar
cluster using the Henry Draper (HD) Catalogue \citep{Cannon24}.
Such stars should be very prominent: at $d = 550$~pc, a ZAMS B0
(B9) star without significant obscuration will have $V \simeq
5$~mag (9~mag). Figure \ref{fig_origin}$a$ shows the spatial
distribution of all known B0-B9 stars (green circles) in a $\sim
40\arcdeg \times 40\arcdeg$ area around the GIRL~G318+32 IRAS loop
(magenta). To discriminate stars likely physically associated with
the loop region from foreground contaminants, we designate stars
inside the loop having photometric distances of $350-700$~pc
derived from placement on the $V$ vs. $V-K_{s}$ diagram with red
$+$ symbols in Figure \ref{fig_origin}$a$.

While there is no indication of a residual cluster at the center
of the $\sim 200$~pc diameter loop, we note the presence of $\ga
6$ B-type stars located only $10-30$~pc northwest of CG~12.
Stellar properties of these B-type stars, which we tentatively
call the HD~120958 group, are listed in Table
\ref{tbl_btype_cluster}. Compared to the Orion Nebular Cluster
(ONC) \citep{Stelzer05}, the HD~120958 group has the same number
of B5-B9 stars (5 late-B stars) but only one star of earlier type
$vs.$ 11 in the ONC where the population extends to O7. These
massive stars may be missing because they have gone supernova over
the past $15-30$~Myr, the lifetime of B1-3 stars
\citep{Hirschi05}. This sequence of supernova shocks could have
elongated the CG~12 cloud into its current cometary structure and
triggered repeated episodes of star formation\footnote{We note
that none of the known radio pulsars from the catalog of
\citet{Manchester05} shown as cyan $\times$ symbols in Figure
\ref{fig_origin}$a$ lies in or near the HD~120958 group.  A pulsar
could be ejected from a massive binary after supernova.}.

The star HD~120958 deserves special attention. Among other B-type
members of the group, it is the closest to CG~12 (only $\sim
10$~pc away providing they lie at the same distance) and is
precisely positioned on the axis of the CG~12 cometary tail
(Figure \ref{fig_origin}$b$). It is also the most luminous star in
the region.  A B5 spectral type was originally assigned in the HD
catalog but later reclassified to B3Vne in the Michigan Catalogue
of \citet{Houk82}. Their footnotes on the star indicate Balmer
line cores in emission and extremely broad lines in absorption,
general observational characteristics of classical Be stars
\citep{Porter03}.

While there is now a consensus that a classical Be star is a
rapidly rotating B-type star that produces a disk in its
equatorial plane, no certain picture of the evolutionary state of
Be phenomenon is available \citep[][and references
therein]{Porter03}. Some fraction of Be stars are produced in
binaries \citep{Gies00}, and Be$+$neutron star (NS) systems
represent the largest sub-class of massive X-ray binaries known
\citep{Coe00,Ziolkowski02}. According to the evolutionary model of
such a binary, the progenitor of the Be star acquires its rapid
rotation via mass and angular momentum transfer from the primary
companion, which in turn undergoes a SN stage and turns into a NS
\citep{vandenHeuvel83}. X-ray emission from Be$+$NS systems is
usually transient.

It is thus reasonable to suggest that HD~120958, well-positioned
along the geometric axis of CG~12, could be the relic of a massive
binary system where the primary has evolved into a SN.  The
encounter of the SN remnant with the cloud could be, at least in
part, responsible for both the star formation and the cometary
morphology of the cloud. We have found no evidence for a transient
X-ray source coincident with HD~120958\footnote{There is a
confused history concerning a report of a transient X-ray source
near the tail of CG~12 with the non-imaging Ariel~V satellite
\citep{Cooke76}. Incorrectly labelled A~0353-40 rather than
A~1353-40, it does not appear in any official Ariel~V source
catalog including the catalog of Ariel~V transients \citep{Pye83}.
Even if the source exists, it lies more than a degree from
HD~120958 and cannot be associated with this star.}.

Another, perhaps less plausible, scenario considering HD~120958 to
be responsible for the morphology of CG~12 and origin of its
stellar population, can be suggested. In this case the B3e
HD~120958 star itself could be the primary ionizing source of the
region. As in the hundred known cases of small bright rimmed
clouds (BRCs) believed to exibit the RDI mechanism, HD~120958
would produce an ionization front at the surface of the CG~12
cloud, which could slowly ablate and drive shocks into the cloud.
In this scenario the poorly studied B77~146 nebula
\citep{Bernes77}, located just at the surface of the cloud
(Figures \ref{fig_dss_combined} and \ref{fig_combined_rgb}$b$) and
interpreted as a reflection nebula powered by the A2 member of the
region \citep{Maheswar04}, could be in fact an ionization front
produced by the UV radiation from HD~120958. However in comparison
with other known BRCs, the CG~12 again would represent a rather
atypical case with a cooler and more distant primary ionizing
source. Several BRCs illuminated by a single B-type are known, but
all with spectral types earlier than B2, and with distances to
BRCs $<4$~pc \citep{Morgan04,Thompson04b}.

\section{Conclusions \label{conclusion_section}}

We report results from a mosaic of four $Chandra$ ACIS-I images
($5 \times 5$~pc$^2$) designed to study the young stellar
population and star formation in the high-latitude cometary
globule CG~12. The main findings of our study are as follows:

1. Of the 128 X-ray sources detected in the $Chandra$ fields
covering the globule's head and its vicinity, 55 have been
identified with known ONIR objects. In addition to several known
young intermediate-mass stars,  we discovered a previously unknown
population of $\ga 50$ T-Tauri members.  Most are unobscured but
some lie inside the cloud.  Three protostars embedded in the
molecular cloud cores are detected, one of which is newly
reported. We establish that the remaining $\sim 70$ X-ray sources
are likely contaminants, mostly extragalactic, based on
simulations of Galactic and extragalactic populations, the
presence or absence of ONIR counterparts, spatial distribution,
and X-ray properties.

2. Based on available ONIR photometry of the CG~12 T-Tauri stars,
we infer they are mostly low-mass $0.2-0.7$~M$_{\odot}$ stars,
$\sim 9-15\%$ of which possess $K_s$ disks. This low fraction of
$K-$excess stars is similar to older T-Tauri populations (e.g.
$\eta$~Cha with age $8-9$~Myr)  and substantially lower than that
of younger PMS populations (e.g. the Taurus clouds).

3. Using the known $L_x-M$ corrections, we combine X-ray and ONIR
properties of the unobscured T-Tauri stars to independently
confirm that the distance to CG~12 must be $>500$~pc, and probably
in the range $550-630$~pc.  This strengthens previous ONIR
findings that CG~12 is indeed located $\ga 200$~pc above the
Galactic plane, and that its cometary morphology with no signs of
nearby hot OB stars is unusual.

4. An evaluation of stellar ages gives a median photometric
isochronal age of $\sim 4$~Myr with a large spread from $<1$~Myr
to $\sim 10-20$~Myr.  The X-ray properties of three deeply
embedded stars confirm the protostellar classification and
previous infrared-radio findings of ongoing star formation in the
globule's head.

5. Assuming the Galactic field IMF for the unobscured YSO
population of CG~12, we estimate a total population of $\sim 80$
stars associated with CG~12 with masses $> 0.2$ M$_\odot$. The
star formation efficiency of the globule is in the range
$15-35$\%, typical for active star forming regions, but above that
seen in smaller molecular globules with star formation triggered
by UV shock fronts.

6. The age and spatial distribution of young stars is consistent
with that seen in other star forming regions, but is inconsistent
with a simple model of radiative driven implosion often applied to
explain CG morphology and star formation.  This suggests either
that CG~12 is atypically large so that a triggering shock takes
millions of years to traverse the cloud, or that CG~12 has been
subject to several distinct episodes of triggering and ablation
over millions of years.

7. Examination of the vicinity of the cloud leads to a revised
scenario for the origin of the CG~12 star forming region. We
identify a group of B-type stars northwest of CG~12 and infer that
a dozen expected early O-B3 type members are missing. These could
be the source of multiple SN explosions triggering star formation
in the cloud over an extended period of $15-30$~Myr. The B3e star
HD~120958 is the closest and most luminous member of the group,
lying 10~pc (projected) from the globule head in the correct
orientation to ablate the cloud into its cometary structure.  It
might itself be the primary ionizing source driving an RDI shock
into the CG~12 cloud, or might be the relic of a massive binary
system that has undergone SN explosion.

\acknowledgements We thank Sangwook Park and Oleg Kargaltsev for
useful discussions on H~I shells and pulsars. This work is
supported by $Chandra$ guest observer grant SAO~G06-7007X (PI:
K.V. Getman). This work was also supported by the $Chandra$ ACIS
Team (G. Garmire, PI) through NASA contract NAS8-38252. This
publication makes use of data products from the Two Micron All Sky
Survey (a joint project of the University of Massachusetts and the
Infrared Processing and Analysis Center/California Institute of
Technology, funded by NASA and NSF). This research makes use of
the SIMBAD database, operated at CDS, Strasbourg, France.

\clearpage
\begin{deluxetable}{cccrccc}
\centering \tabletypesize{\small} \tablewidth{0pt}
\tablecolumns{7}

\tablecaption{CG~12 Observations
\label{observation_journal_table}}

\tablehead{

\colhead{Field} & \colhead{ObsID} & \colhead{Start Time} &
\colhead{Exposure} & \colhead{R.A. (J2000.0)} & \colhead{Decl.
(J2000.0)}
& \colhead{Roll Angle} \\

&&& (ks) & (hh:mm:ss.ss) & (dd:mm:ss.ss) & (deg) }

\startdata
I & 6423 & 2006 Apr 15 16:19:17 & 30.8 & 13:57:44.52 & -39:58:48.31 & 11.5 \\
II & 6424 & 2006 Jun 02 07:25:09 & 3.1 & 13:57:42.87 & -39:43:01.76 & 285.0 \\
III & 6425 & 2006 Apr 13 08:44:08 & 3.1 & 13:56:19.40 & -39:42:47.94 & 14.7 \\
IV & 6426 & 2006 Apr 15 12:54:20 & 3.1 & 13:56:19.40 & -39:58:48.09 & 11.1 \\
\enddata

\tablecomments{ObsID values are from the $Chandra$ Observation
Catalog.  Start Times are in UT.  Exposure times are the sum of
Good Time Intervals (GTIs) for the CCD at the telescope aim point
(CCD3) minus 1.3\% to account for CCD readouts. The aimpoints and
roll angles are obtained from the satellite aspect solution before
astrometric correction was applied.}

\end{deluxetable}

\clearpage
\begin{deluxetable}{crcrrrrrrrrrrrcccccc}
\centering \rotate \tabletypesize{\tiny} \tablewidth{0pt}
\tablecolumns{20}

\tablecaption{Basic Source Properties \label{tbl_bsp}}

\tablehead{ \multicolumn{3}{c}{Source} &
  &
\multicolumn{4}{c}{Position} &
  &
\multicolumn{5}{c}{Extracted Counts} &
  &
\multicolumn{5}{c}{Characteristics} \\
\cline{1-3} \cline{5-8} \cline{10-14} \cline{16-20}

\colhead{Field} & \colhead{Seq} & \colhead{CXOU~J} &
  &
\colhead{$\alpha_{\rm J2000}$} & \colhead{$\delta_{\rm J2000}$} &
\colhead{Err} & \colhead{$\theta$} &
  &
\colhead{Net} & \colhead{$\Delta$Net} & \colhead{Bkgd} &
\colhead{Net} & \colhead{PSF} &
  &
\colhead{Signif} & \colhead{Anom\tablenotemark{a}} & \colhead{Var\tablenotemark{b}} & \colhead{EffExp} & \colhead{Med E}\\

\colhead{} & \colhead{\#} & \colhead{} &
  &
\colhead{(deg)} & \colhead{(deg)} & \colhead{(\arcsec)} &
\colhead{(\arcmin)} &
  &
\colhead{Full} & \colhead{Full} & \colhead{Full} & \colhead{Hard}
& \colhead{Frac} &
  &
\colhead{} & \colhead{} & \colhead{} & \colhead{(ks)} &
\colhead{(keV)}\\

\colhead{(1)} & \colhead{(2)} & \colhead{(3)} &
  &
\colhead{(4)} & \colhead{(5)} & \colhead{(6)} & \colhead{(7)} &
  &
\colhead{(8)} & \colhead{(9)} & \colhead{(10)} & \colhead{(11)} &
\colhead{(12)} &
  &
\colhead{(13)} & \colhead{(14)} & \colhead{(15)} & \colhead{(16)}
& \colhead{(17)} }

\startdata
I & 1 & 135657.01$-$395349.7 & & 209.237580 & -39.897144 & 1.25 & 10.38 & & 8.2 & 5.0 & 10.8 & 6.3 & 0.91 & & 1.5 & .... & a & 20.1 & 2.23 \\
I & 2 & 135702.10$-$395409.6 & & 209.258790 & -39.902694 & 0.67 & 9.37 & & 34.4 & 6.9 & 6.6 & 21.9 & 0.89 & & 4.6 & .... & a & 21.2 & 2.61 \\
I & 3 & 135703.91$-$395455.0 & & 209.266310 & -39.915279 & 1.01 & 8.70 & & 8.4 & 4.3 & 5.6 & 7.5 & 0.90 & & 1.7 & .... & a & 22.5 & 3.73 \\
I & 4 & 135704.65$-$400123.2 & & 209.269380 & -40.023117 & 1.24 & 8.06 & & 4.5 & 3.4 & 3.5 & 2.5 & 0.90 & & 1.1 & g... & \nodata & 22.2 & 3.45 \\
I & 5 & 135706.07$-$395822.2 & & 209.275330 & -39.972848 & 0.35 & 7.38 & & 64.2 & 8.7 & 2.8 & 22.3 & 0.90 & & 7.0 & .... & a & 24.6 & 1.61 \\
I & 6 & 135708.01$-$395821.9 & & 209.283380 & -39.972767 & 0.65 & 7.01 & & 13.1 & 4.5 & 2.9 & 8.1 & 0.90 & & 2.6 & .... & a & 25.0 & 3.04 \\
I & 7 & 135709.47$-$400142.7 & & 209.289470 & -40.028538 & 0.44 & 7.32 & & 39.4 & 7.0 & 2.6 & 9.4 & 0.89 & & 5.2 & .... & a & 22.7 & 1.47 \\
I & 8 & 135709.55$-$400528.4 & & 209.289810 & -40.091237 & 1.21 & 9.45 & & 6.5 & 4.2 & 6.5 & 1.9 & 0.89 & & 1.4 & .... & a & 20.6 & 3.79 \\
\enddata

\tablecomments{Column 1: $Chandra$ ACIS-I fields surveyed in the
CG~12 region (I - primary pointing; II, III, and IV - secondary
pointings). Column 2: X-ray source number. Column 3: IAU
designation. Columns 4-5: Right ascension and declination for
epoch J2000.0 in degrees. Column 6: Positional uncertainty. Column
7: Off-axis angle. Columns 8-9: Estimated net counts from
extracted area in total energy band $(0.5-8.0)$ keV and their
1\/$\sigma$ errors. Column 10: Estimated background counts from
extracted area in total energy band. Column 11: Estimated net
counts from extracted area in hard energy band $(2-8)$ keV. Column
12: Fraction of the PSF at the fiducial energy of 1.497 keV
enclosed within the extracted area. Note that a reduced PSF
fraction (significantly below 90\%) may indicate that the source
is in a crowded region or on the edge of the field. Column 13:
Source significance in sigma. Column 14. Source anomalies
described by a set of flags, see the text note$^{\rm{a}}$ below
for details. Column 15: Variability characterization based on K-S
statistics, see the text note$^{\rm{b}}$ below for details. Column
16: Source effective exposure in ks. Column 17: The
background-corrected median photon energy in the total $0.5-8$ keV
energy band.}

\tablenotetext{a}{ Source anomalies:  g = fractional time that
source was on a detector (FRACEXPO from {\em mkarf}) is $<0.9$ ; e
= source on field edge; p = source piled up; s = source on readout
streak.}

\tablenotetext{b}{ Source variability:  a = no evidence for
variability; b = possibly variable; c = definitely variable. No
test is performed for sources with fewer than 4 total full-band
counts. No value is reported for sources in chip gaps or on field
edges.}

\end{deluxetable}

\clearpage

\begin{deluxetable}{rrcrrccccrrrr}
\centering \rotate \tabletypesize{\scriptsize} \tablewidth{0pt}
\tablecolumns{13}

\tablecaption{X-ray Spectroscopy for Photometrically Selected
Sources:  Thermal Plasma Fits \label{tbl_thermal_spectroscopy}}

\tablehead{ \multicolumn{5}{c}{Source} &
  &
\multicolumn{2}{c}{Spectral Fit\tablenotemark{a}} &
  &
\multicolumn{4}{c}{X-ray Luminosities\tablenotemark{b}} \\
\cline{1-5} \cline{7-8} \cline{10-13}

\colhead{Field} & \colhead{Seq} & \colhead{CXOU J} & \colhead{Net}
& \colhead{Med E} &
  &
\colhead{$\log N_H$} & \colhead{$kT$} &
  &
\colhead{$\log L_h$} & \colhead{$\log L_{h,c}$} & \colhead{$\log
L_t$} & \colhead{$\log L_{t,c}$} \\

\colhead{} & \colhead{\#} & \colhead{} & \colhead{(counts)} &
\colhead{(keV)} &
  &
\colhead{(cm$^{-2}$)} & \colhead{(keV)} &
  &
\multicolumn{4}{c}{(ergs s$^{-1}$)} \\

\colhead{(1)} & \colhead{(2)} & \colhead{(3)} & \colhead{(4)} &
\colhead{(5)} &
  &
\colhead{(6)} & \colhead{(7)} &
  &
\colhead{(8)} & \colhead{(9)} & \colhead{(10)} &\colhead{(11)} }

\startdata
I & 2 & 135702.10$-$395409.6 &    34.4 &  2.61 &  &  {\tiny $-1.0$} 21.7 {\tiny $+0.5$} & {\tiny $-11.7$} 13.0 \phantom{{\tiny $-11.7$}}         &  & 29.71 & 29.73 & 29.80 & 29.90\\
I & 5 & 135706.07$-$395822.2 &    64.2 &  1.61 &  &   20.0  &                                                 11.7                                &  & 29.82 & 29.82 & 29.99 & 29.99\\
I & 7 & 135709.47$-$400142.7 &    39.4 &  1.47 &  &   21.5  &  2.5                                                                                &  & 29.49 & 29.51 & 29.72 & 29.88\\
I & 9 & 135711.34$-$400234.7 &   201.1 &  1.10 &  &   20.0  &  1.2                                                                                &  & 29.49 & 29.49 & 30.26 & 30.26\\
I & 10 & 135712.03$-$400050.2 &    41.8 &  1.16 &  &   20.0 &  1.2                                                                                &  & 29.08 & 29.08 & 29.82 & 29.83\\
I & 11 & 135712.31$-$400026.5 &    22.3 &  1.55 &  &  \phantom{{\tiny $+0.3$}} 21.6 {\tiny $+0.3$} & {\tiny $-1.4$} 2.6 \phantom{{\tiny $-1.4$}} &  & 29.25 & 29.27 & 29.45 & 29.63\\
I & 14 & 135716.75$-$395234.2 &    50.6 &  1.50 &  &   21.6  &  4.1                                                                               &  & 29.76 & 29.78 & 29.91 & 30.05\\
I & 17 & 135717.06$-$395710.5 &    25.1 &  1.00 &  &   20.0  &  0.8                                                                               &  & 28.21 & 28.21 & 29.36 & 29.37\\
\enddata

\tablecomments{Columns 1-5: For convenience, these columns
reproduce the source identification, photometric net counts, and
median energy data from Table \ref{tbl_bsp}. Columns 6-7: Best-fit
values for the column density and plasma temperature parameters.
See the text note$^{\rm{a}}$ below for details. Columns 8-11:
X-ray luminosities obtained from spectral analysis. See the text
note$^{\rm{b}}$ below for details.}

\tablenotetext{a}{All fits used the ``wabs(apec)'' model in {\it
XSPEC} and assumed 0.3$Z_{\odot}$ abundances. Uncertainties
represent 90\% confidence intervals. More significant digits are
used for uncertainties $<0.1$ in order to avoid large rounding
errors; for consistency, the same number of significant digits is
used for both lower and upper uncertainties. Uncertainties are
missing when {\it XSPEC} was unable to compute them or when their
values were so large that the parameter is effectively
unconstrained. Fitted values with $\log N_H < 20.0$~cm$^{-2}$ are
truncated at 20.0 because ACIS-I spectra are insensitive to
differences in very low column densities. Fitted values with $kT >
15$~keV are truncated at 15~keV because the data often cannot
discriminate between very high-temperature values.}

\tablenotetext{b}{X-ray luminosities are calculated assuming a
distance of 550~pc. h = hard band (2--8 keV); t = total band
(0.5--8 keV). Absorption-corrected luminosities are subscripted
with a $c$.}

\end{deluxetable}

\clearpage

\begin{deluxetable}{crccrcrrrrcrrrrc}
\centering \rotate \tabletypesize{\tiny} \tablewidth{0pt}
\tablecolumns{16}

\tablecaption{Optical and NIR Counterparts \label{tbl_cntrpts}}

\tablehead{ \colhead{Field} & \colhead{Seq} & \colhead{CXOU~J} &
\colhead{NOMAD} & \colhead{Sep1} & \colhead{Mah04} &
\colhead{$V$\tablenotemark{a}} & \colhead{$R$\tablenotemark{a}} &
\colhead{$\mu_{\alpha\ast}$} & \colhead{$\mu_{\delta}$}
& \colhead{2MASS} & \colhead{Sep2} & \colhead{$J$} & \colhead{$H$} & \colhead{$K_s$} & \colhead{Fl}\\

\colhead{} & \colhead{\#} & \colhead{} & \colhead{} &
\colhead{(\arcsec)} & \colhead{} & \colhead{(mag)} &
\colhead{(mag)} & \colhead{(mas/yr)} & \colhead{(mas/yr)} &
\colhead{} & \colhead{(\arcsec)} & \colhead{(mag)} &
\colhead{(mag)} & \colhead{(mag)} & \colhead{}\\

\colhead{(1)} & \colhead{(2)} & \colhead{(3)} & \colhead{(4)} &
\colhead{(5)} & \colhead{(6)} & \colhead{(7)} & \colhead{(8)} &
\colhead{(9)} & \colhead{(10)} & \colhead{(11)} & \colhead{(12)} &
\colhead{(13)} & \colhead{(14)} & \colhead{(15)} & \colhead{(16)}
}

\startdata
I & 1 & 135657.01-395349.7 & \nodata & \nodata & \nodata & \nodata & \nodata & \nodata & \nodata & \nodata & \nodata & \nodata & \nodata & \nodata & \nodata\\
I & 2 & 135702.10-395409.6 & \nodata & \nodata & \nodata & \nodata & \nodata & \nodata & \nodata & \nodata & \nodata & \nodata & \nodata & \nodata & \nodata\\
I & 3 & 135703.91-395455.0 & \nodata & \nodata & \nodata & \nodata & \nodata & \nodata & \nodata & \nodata & \nodata & \nodata & \nodata & \nodata & \nodata\\
I & 4 & 135704.65-400123.2 & \nodata & \nodata & \nodata & \nodata & \nodata & \nodata & \nodata & \nodata & \nodata & \nodata & \nodata & \nodata & \nodata\\
I & 5 & 135706.07-395822.2 & 0500-0310911 &   0.4 & \nodata & \nodata &  18.72 & \nodata & \nodata & \nodata & \nodata & \nodata & \nodata & \nodata & \nodata\\
I & 6 & 135708.01-395821.9 & \nodata & \nodata & \nodata & \nodata & \nodata & \nodata & \nodata & \nodata & \nodata & \nodata & \nodata & \nodata & \nodata\\
I & 7 & 135709.47-400142.7 & \nodata & \nodata & \nodata & \nodata & \nodata & \nodata & \nodata & \nodata & \nodata & \nodata & \nodata & \nodata & \nodata\\
I & 8 & 135709.55-400528.4 & \nodata & \nodata & \nodata & \nodata & \nodata & \nodata & \nodata & \nodata & \nodata & \nodata & \nodata & \nodata & \nodata\\
\enddata

\tablecomments{Column 1: $Chandra$ ACIS-I fields surveyed in the
CG~12 region (I - primary pointing; II, III, and IV - secondary
pointings). Column 2: X-ray source number. Column 3: IAU
designation. Column 4: NOMAD identifier. Column 5: $Chandra$-NOMAD
positional offset. Column 6: Source identifier from
\citet{Maheswar04} in the format: star number - region number.
Columns 7-8: $V$ and $R$ magnitudes, see the text note$^{\rm{a}}$
below for details. Columns 9-10: ``Best'' proper motions and their
mean errors from catalogs compiled in NOMAD. $\mu_{\alpha\ast} =
\mu_{\alpha} \times \cos (\delta)$. Column 11: 2MASS name. Column
12: $Chandra$-2MASS positional offset. Columns 13-15: 2MASS
$JHK_s$ magnitudes and their formal statistical errors. Column 16:
2MASS photometry quality and confusion-contamination flag.}

\tablenotetext{a}{When formal statistical errors are given, $V$
and $R$ magnitudes are from \citet{Maheswar04}. $V$ magnitudes for
3 B-type stars, I$_{-}57$, I$_{-}58$, and I$_{-}74$, are from
\citet{Williams77}. Otherwise $V$ and $R$ magnitudes are from
NOMAD catalog, in which case all $V$ magnitudes except for the
$Chandra$ source I$_{-}81$ are from YB6 (= USNO unpublished)
catalog, and all $R$ magnitudes except for the $Chandra$ source
I$_{-}57$ are from USNO-B1.0 catalog ($R$ magnitudes from
USNO-B1.0 are second red magnitudes $R_2$, except for the
$Chandra$ sources I$_{-}70$ and II$_{-}9$, for which it is a first
red magnitude $R_1$). For the $Chandra$ source I$_{-}81$ $V$
magnitude is from Tycho-2 catalog, and for the $Chandra$ source
I$_{-}57$ $R$ magnitude is from UCAC2 catalog.}

\end{deluxetable}

\clearpage
\begin{deluxetable}{ccccccc}
\centering \tabletypesize{\small} \tablewidth{0pt}
\tablecolumns{7}

\tablecaption{B-type members of the stellar group northwest of
CG~12 \label{tbl_btype_cluster}}

\tablehead{

\colhead{ID} & \colhead{R.A. (J2000.0)} & \colhead{Decl.
(J2000.0)} & \colhead{Sp. Type} & \colhead{$V$} & \colhead{$K_s$}
& \colhead{$\pi$} \\

&(hh:mm:ss.ss)&(dd:mm:ss.ss)&  & (mag) & (mag)  & (mas)\\

(1)&(2)&(3)&(4)&(5)&(6)&(7)}

\startdata
HD~120958 & 13:53:28.228 & -39:03:25.92 & B3Vne & 7.61 & 7.613 & $2.20 \pm 0.85$ \\
HD~120124 & 13:48:15.113 & -38:34:12.43 & B8V   & 9.19 & 9.384 & \nodata \\
HD~119627 & 13:45:00.702 & -38:28:52.39 & B7IV/V   & 8.94 & 9.006 & \nodata \\
HD~119484 & 13:44:13.497 & -38:46:29.62 & B8IV/V  & 9.38 & 9.332 & \nodata \\
HD~119338 & 13:43:25.294 & -38:15:23.55 & B5V  & 8.91 & 9.177 & $1.76 \pm 1.01$ \\
HD~119277 & 13:43:00.718 & -38:20:08.45 & B9  & 8.81 & 9.053 & \nodata \\
\enddata

\tablecomments{Column 1: Source identification from HD catalog.
Columns 2-5: Source position, spectral type, and $V$ magnitude
from SIMBAD. Column 6: $K_s$ magnitude from 2MASS catalog. Column
7. Parallax from SIMBAD.}

\end{deluxetable}

\clearpage

\clearpage

\clearpage

\clearpage

\clearpage

\centerline{\includegraphics[angle=0.,width=6.5in]{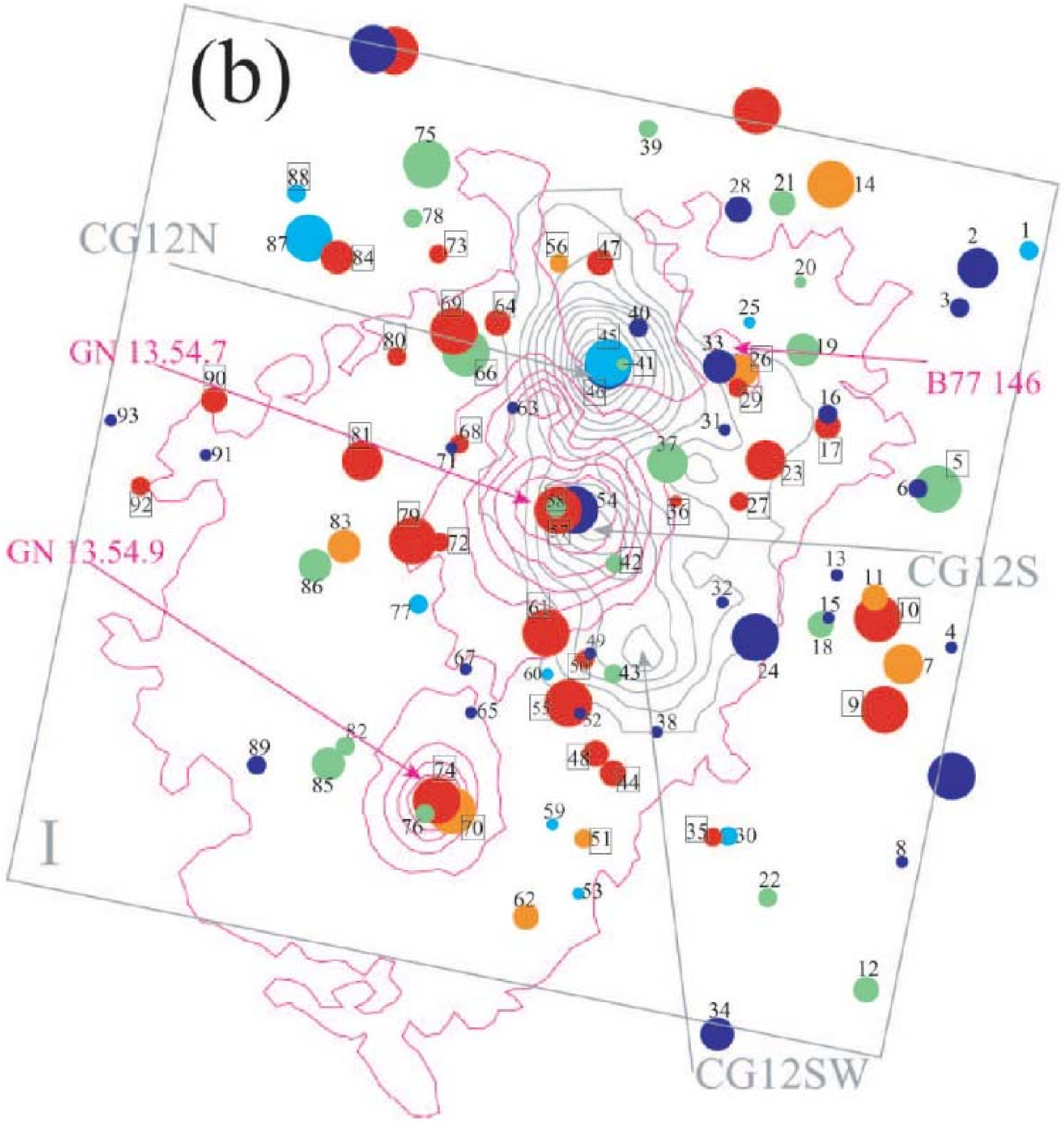}}

\clearpage

\clearpage

\clearpage

\clearpage

\clearpage

\clearpage

\includegraphics[angle=0.,width=6.5in]{f9b.eps}

\clearpage

\clearpage

\clearpage

\clearpage

\centerline{\includegraphics[angle=0.,width=6.5in]{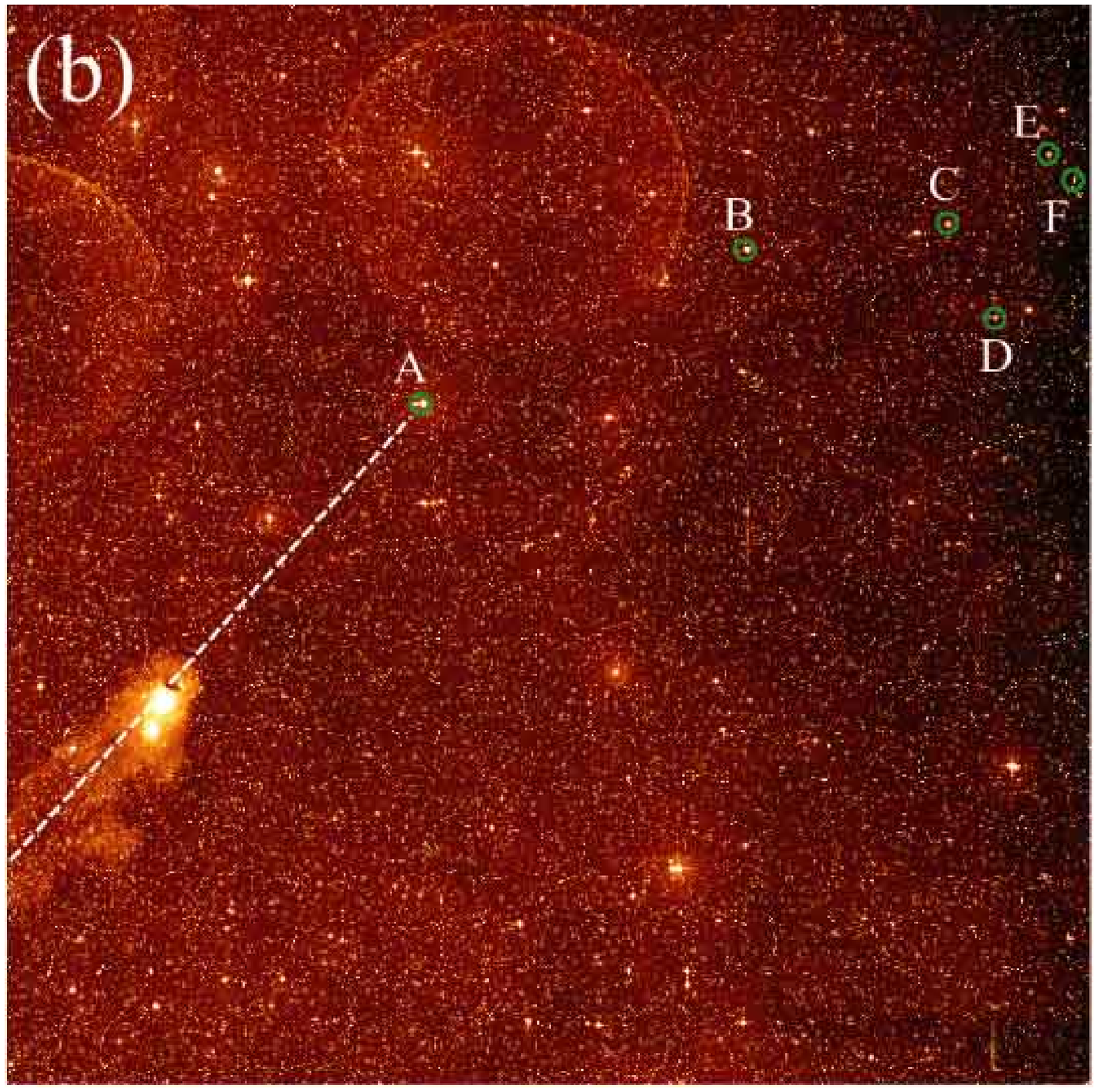}}

\clearpage

\clearpage

\begin{thebibliography}



\bibitem[Anders \& Grevesse(1989)]{Anders89} Anders, E., \&
Grevesse, N.\ 1989, \gca, 53, 197



\bibitem[Argiroffi et al.(2006)]{Argiroffi06} Argiroffi, C.,
Favata, F., Flaccomio, E., Maggio, A., Micela, G., Peres, G., \&
Sciortino, S.\ 2006, \aap, 459, 199



\bibitem[Arnaud(1996)]{Arnaud96} Arnaud, K.\ A.\ 1996, in Data
Analysis Software and Systems V, ed. G.\ H.\ Jacoby \& J.\ Barnes
(San Francisco:ASP), 17








\bibitem[Ballesteros-Paredes \& Hartmann(2007)]{Ballesteros-Paredes07}
Ballesteros-Paredes, J., \& Hartmann, L.\ 2007, Revista Mexicana
de Astronomia y Astrofisica, 43, 123



\bibitem[Baraffe et al.(1998)]{Baraffe98} Baraffe, I., Chabrier,
G., Allard, F., \& Hauschildt, P.~H.\ 1998, \aap, 337, 403



\bibitem[Barger et al.(2003)]{Barger03} Barger, A.~J., et al.\
2003, \aj, 126, 632


\bibitem[Bate et al.(2003)]{Bate03} Bate, M.~R., Bonnell,
I.~A., \& Bromm, V.\ 2003, \mnras, 339, 577



\bibitem[Bauer et al.(2003)]{Bauer03} Bauer, F.~E., et al.\
2003, Astronomische Nachrichten, 324, 175



\bibitem[Bernes(1977)]{Bernes77} Bernes, C.\ 1977, \aaps, 29, 65


\bibitem[Bertoldi(1989)]{Bertoldi89} Bertoldi, F.\ 1989, \apj,
346, 735



\bibitem[Bessell \& Brett(1988)]{Bessell88} Bessell, M.~S., \&
Brett, J.~M.\ 1988, \pasp, 100, 1134



\bibitem[Bok \& Reilly(1947)]{Bok47} Bok, B.~J., \& Reilly,
E.~F.\ 1947, \apj, 105, 255



\bibitem[Bourke et al.(1995a)]{Bourke95a} Bourke, T.~L., Hyland,
A.~R., \& Robinson, G.\ 1995a, \mnras, 276, 1052



\bibitem[Bourke et al.(1995b)]{Bourke95b} Bourke, T.~L., Hyland,
A.~R., Robinson, G., James, S.~D., \& Wright, C.~M.\ 1995b,
\mnras, 276, 1067



\bibitem[Bouwman et al.(2006)]{Bouwman06} Bouwman, J., Lawson,
W.~A., Dominik, C., Feigelson, E.~D., Henning, T., Tielens,
A.~G.~G.~M., \& Waters, L.~B.~F.~M.\ 2006, \apjl, 653, L57



\bibitem[Brandt et al.(2001)]{Brandt01} Brandt, W.~N., et al.\
2001, \aj, 122, 2810



\bibitem[Burningham et al.(2005)]{Burningham05} Burningham, B.,
Naylor, T., Littlefair, S.~P., \& Jeffries, R.~D.\ 2005, \mnras,
363, 1389



\bibitem[Cannon \& Pickering(1924)]{Cannon24} Cannon, A.~J., \&
Pickering, E.~C.\ 1924, Annals of Harvard College Observatory, 99,
1



\bibitem[Carpenter et al.(2001)]{Carpenter01} Carpenter, J.~M.,
Hillenbrand, L.~A., \& Skrutskie, M.~F.\ 2001, \aj, 121, 3160



\bibitem[Cash(1979)]{Cash79} Cash, W.\ 1979, \apj, 228, 939



\bibitem[Clarke et al.(2000)]{Clarke00} Clarke, C.~J., Bonnell,
I.~A., \& Hillenbrand, L.~A.\ 2000, Protostars and Planets IV, 151



\bibitem[Coe(2000)]{Coe00} Coe, M.~J.\ 2000, IAU Colloq.~175:
The Be Phenomenon in Early-Type Stars, 214, 656



\bibitem[Cooke(1976)]{Cooke76} Cooke, B.~A.\ 1976, \nat, 261,
564


\bibitem[Cutri et al.(2003)]{Cutri03} Cutri, R.~M., et al.\
2003, 2MASS All Sky Catalog of point sources, published by The
IRSA 2MASS All-Sky Point Source Catalog, NASA/IPAC Infrared
Science Archive.~http://irsa.ipac.caltech.edu/applications/Gator/



\bibitem[Deharveng et al.(2005)]{Deharveng05} Deharveng, L.,
Zavagno, A., \& Caplan, J.\ 2005, \aap, 433, 565



\bibitem[De Vries et al.(2002)]{deVries02} De Vries, C.~H.,
Narayanan, G., \& Snell, R.~L.\ 2002, \apj, 577, 798



\bibitem[Dickey \& Lockman(1990)]{Dickey90} Dickey, J.~M., \&
Lockman, F.~J.\ 1990, \araa, 28, 215



\bibitem[Dutra \& Bica(2002)]{Dutra02} Dutra, C.~M., \& Bica,
E.\ 2002, \aap, 383, 631



\bibitem[Ehlerov{\'a} \& Palou{\v s}(2005)]{Ehlerova05}
Ehlerov{\'a}, S., \& Palou{\v s}, J.\ 2005, \aap, 437, 101



\bibitem[Elmegreen et al.(2000)]{Elmegreen00} Elmegreen, B.~G.,
Efremov, Y., Pudritz, R.~E., \& Zinnecker, H.\ 2000, Protostars
and Planets IV, 179



\bibitem[Elmegreen(2000b)]{Elmegreen00b} Elmegreen, B.~G.\ 2000b,
\apj, 530, 277



\bibitem[Feigelson(1996)]{Feigelson96} Feigelson, E.~D.\ 1996,
\apj, 468, 306



\bibitem[Feigelson et al.(2002)]{Feigelson02} Feigelson, E.~D.,
Broos, P., Gaffney, J.~A., Garmire, G., Hillenbrand, L.~A.,
Pravdo, S.~H., Townsley, L., \& Tsuboi, Y.\ 2002, \apj, 574, 258



\bibitem[Feigelson et al.(2003)]{Feigelson03} Feigelson, E.~D.,
Lawson, W.~A., \& Garmire, G.~P.\ 2003, \apj, 599, 1207



\bibitem[Feigelson et al.(2005)]{Feigelson05} Feigelson, E.~D., et
al.\ 2005, \apjs, 160, 379


\bibitem[Feigelson et al.(2007)]{Feigelson07} Feigelson, E.,
Townsley, L., G{\"u}del, M., \& Stassun, K.\ 2007, Protostars and
Planets V, 313


\bibitem[Garmire et al.(2003)]{Garmire03} Garmire, G.~P., Bautz,
M.~W., Ford, P.~G., Nousek, J.~A., \& Ricker, G.~R.\ 2003,
\procspie, 4851, 28



\bibitem[Gehrels(1986)]{Gehrels86} Gehrels, N.\ 1986, \apj, 303,
336



\bibitem[Getman et al.(2002)]{Getman02} Getman, K.~V.,
Feigelson, E.~D., Townsley, L., Bally, J., Lada, C.~J., \&
Reipurth, B.\ 2002, \apj, 575, 354



\bibitem[Getman et al.(2005a)]{Getman05a} Getman, K.~V., Flaccomio,
E., Broos, P. et al.\ 2005a, \apjs, 160, 319



\bibitem[Getman et al.(2005b)]{Getman05b} Getman, K.~V.,
Feigelson, E.~D., Grosso, N., McCaughrean, M.~J., Micela, G.,
Broos, P., Garmire, G., \& Townsley, L.\ 2005b, \apjs, 160, 353



\bibitem[Getman et al.(2006)]{Getman06} Getman, K.~V.,
Feigelson, E.~D., Townsley, L., Broos, P., Garmire, G., \&
Tsujimoto, M.\ 2006, \apjs, 163, 306



\bibitem[Getman et al.(2007)]{Getman07} Getman, K.~V.,
Feigelson, E.~D., Garmire, G., Broos, P., \& Wang, J.\ 2007, \apj,
654, 316


\bibitem[Giardino et al.(2007)]{Giardino07} Giardino, G., Favata,
F., Micela, G., Sciortino, S., \& Winston, E.\ 2007, \aap, 463,
275


\bibitem[Gies(2000)]{Gies00} Gies, D.~R.\ 2000, IAU
Colloq.~175: The Be Phenomenon in Early-Type Stars, 214, 668


\bibitem[Gorti \& Hollenbach(2002)]{Gorti02} Gorti, U., \&
Hollenbach, D.\ 2002, \apj, 573, 215



\bibitem[Guedel et al.(2006)]{Guedel06} Guedel, M., et al.\
2006, \aap, in press (astro-ph/0609160)



\bibitem[Haikala \& Olberg(2006)]{Haikala06} Haikala, L.~K., \&
Olberg, M.\ 2006, \aap, in press (astro-ph/0609820)



\bibitem[Haisch et al.(2001)]{Haisch01} Haisch, K.~E., Jr.,
Lada, E.~A., \& Lada, C.~J.\ 2001, \apjl, 553, L153



\bibitem[Hartmann et al.(2001)]{Hartmann01} Hartmann, L.,
Ballesteros-Paredes, J., \& Bergin, E.~A.\ 2001, \apj, 562, 852


\bibitem[Hartmann(2003)]{Hartmann03} Hartmann, L.\ 2003, \apj,
585, 398


\bibitem[Hawarden \& Brand(1976)]{Hawarden76} Hawarden, T.~G., \&
Brand, P.~W.~J.~L.\ 1976, \mnras, 175, 19P



\bibitem[Heiles(1984)]{Heiles84} Heiles, C.\ 1984, \apjs, 55,
585



\bibitem[Herbst et al.(2002)]{Herbst02} Herbst, W.,
Bailer-Jones, C.~A.~L., Mundt, R., Meisenheimer, K., \&
Wackermann, R.\ 2002, \aap, 396, 513



\bibitem[Herschel(1847)]{Herschel1847} Herschel, J.~F.~W.~S.\ 1847,
London, Smith, Elder and co., 1847



\bibitem[Hirschi et al.(2005)]{Hirschi05} Hirschi, R., Meynet,
G., \& Maeder, A.\ 2005, \aap, 433, 1013


\bibitem[Houk \& Fuentes-Williams(1982)]{Houk82} Houk, N., \&
Fuentes-Williams, T.~H.\ 1982, \baas, 14, 615



\bibitem[H{\" u}nsch et al.(1999)]{Hunsch99} H{\" u}nsch, M.,
Schmitt, J.~H.~M.~M., Sterzik, M.~F., \& Voges, W.\ 1999, \aaps,
135, 319



\bibitem[Imanishi, Koyama, \& Tsuboi(2001)]{Imanishi01} Imanishi,
K., Koyama, K., \& Tsuboi, Y.\ 2001, \apj, 557, 747



\bibitem[Jeffries(2007)]{Jeffries07} Jeffries, R.~D.\ 2007,
\mnras, 159



\bibitem[Johnson(1966)]{Johnson66} Johnson, H.~L.\ 1966, \araa,
4, 193



\bibitem[Kessel-Deynet \& Burkert(2003)]{Kessel03}
Kessel-Deynet, O., \& Burkert, A.\ 2003, \mnras, 338, 545



\bibitem[Keto \& Myers(1986)]{Keto86} Keto, E.~R., \& Myers, P.~C.\
1986, \apj, 304, 466



\bibitem[K{\"o}nyves et al.(2007)]{Konyves07} K{\"o}nyves, V.,
Kiss, C., Mo{\'o}r, A., Kiss, Z.~T., \& T{\'o}th, L.~V.\ 2007,
\aap, 463, 1227



\bibitem[Krumholz \& Tan(2007)]{Krumholz07} Krumholz, M.~R., \&
Tan, J.~C.\ 2007, \apj, 654, 304


\bibitem[Larson(1981)]{Larson81} Larson, R.~B.\ 1981, \mnras,
194, 809



\bibitem[Lawson et al.(2001)]{Lawson01} Lawson, W.~A., Crause,
L.~A., Mamajek, E.~E., \& Feigelson, E.~D.\ 2001, \mnras, 321, 57



\bibitem[Lefloch \& Lazareff(1994)]{Lefloch94} Lefloch, B., \&
Lazareff, B.\ 1994, \aap, 289, 559



\bibitem[Lefloch \& Lazareff(1995)]{Lefloch95} Lefloch, B., \&
Lazareff, B.\ 1995, \aap, 301, 522



\bibitem[Lehmann et al.(2001)]{Lehmann01} Lehmann, I., et al.\
2001, \aap, 371, 833



\bibitem[Luhman \& Rieke(1998)]{Luhman98} Luhman, K.~L., \&
Rieke, G.~H.\ 1998, \apj, 497, 354



\bibitem[Luhman \& Steeghs(2004)]{Luhman04} Luhman, K.~L., \&
Steeghs, D.\ 2004, \apj, 609, 917



\bibitem[Lyo et al.(2003)]{Lyo03} Lyo, A.-R., Lawson, W.~A.,
Mamajek, E.~E., Feigelson, E.~D., Sung, E.-C., \& Crause, L.~A.\
2003, \mnras, 338, 616



\bibitem[Lyons(1991)]{Lyons91} Lyons, L.\ 1991, A Practical Guide to Data Analysis for Physical Science Students (Cambridge: Cambridge Univ. Press)


\bibitem[Kenyon \& Hartmann(1995)]{Kenyon95} Kenyon, S.~J., \&
Hartmann, L.\ 1995, \apjs, 101, 117


\bibitem[Kroupa(2002)]{Kroupa02} Kroupa, P.\ 2002, Science, 295,
82



\bibitem[Magakian(2003)]{Magakian03} Magakian, T. Yu.\ 2003, \aap, 399, 141



\bibitem[Maheswar et al.(2004)]{Maheswar04} Maheswar, G., Manoj,
P., \& Bhatt, H.~C.\ 2004, \mnras, 355, 1272



\bibitem[Mamajek et al.(2000)]{Mamajek00} Mamajek, E.~E., Lawson,
W.~A., \& Feigelson, E.~D.\ 2000, \apj, 544, 356



\bibitem[Mamajek et al.(2002)]{Mamajek02} Mamajek, E.~E., Meyer,
M.~R., \& Liebert, J.\ 2002, \aj, 124, 1670


\bibitem[Manchester et al.(2005)]{Manchester05} Manchester, R.~N.,
Hobbs, G.~B., Teoh, A., \& Hobbs, M.\ 2005, \aj, 129, 1993



\bibitem[Marraco \& Forte(1978)]{Marraco78} Marraco, H.~G., \&
Forte, J.~C.\ 1978, \apj, 224, 473



\bibitem[Matsuyanagi et al.(2006)]{Matsuyanagi06} Matsuyanagi, I.,
Itoh, Y., Sugitani, K., Oasa, Y., Mukai, T., \& Tamura, M.\ 2006,
\pasj, 58, L29



\bibitem[Megeath et al.(2005)]{Megeath05} Megeath, S.~T.,
Hartmann, L., Luhman, K.~L., \& Fazio, G.~G.\ 2005, \apjl, 634,
L113



\bibitem[Meyer et al.(1997)]{Meyer97} Meyer, M.~R., Calvet, N.,
\& Hillenbrand, L.~A.\ 1997, \aj, 114, 288



\bibitem[Miao et al.(2006)]{Miao06} Miao, J., White, G.~J.,
Nelson, R., Thompson, M., \& Morgan, L.\ 2006, \mnras, 369, 143



\bibitem[Monet et al.(2003)]{Monet03} Monet, D.~G., et al.\
2003, \aj, 125, 984



\bibitem[Moretti et al.(2003)]{Moretti03} Moretti, A., Campana, S., Lazzati,
D., \& Tagliaferri, G.\ 2003, \apj, 588, 696


\bibitem[Morgan et al.(2004)]{Morgan04} Morgan, L.~K., Thompson,
M.~A., Urquhart, J.~S., White, G.~J., \& Miao, J.\ 2004, \aap,
426, 535




\bibitem[Morrison \& McCammon(1983)]{Morrison83} Morrison, R.~\&
McCammon, D.\ 1983, \apj, 270, 119



\bibitem[Mouschovias et al.(2006)]{Mouschovias06} Mouschovias, T.~C.,
Tassis, K., \& Kunz, M.~W.\ 2006, \apj, 646, 1043


\bibitem[Nakamura \& Li(2007)]{Nakamura07} Nakamura, F., \& Li,
Z.-Y.\ 2007, ArXiv Astrophysics e-prints, astro-ph/0703152


\bibitem[Neckel \& Vehrenberg(1990)]{Neckel90} Neckel, T., \& Vehrenberg, H.\ 1990, Duesseldorf: Treugesell-Verlag, 1990


\bibitem[Odenwald et al.(1992)]{Odenwald92} Odenwald, S., Fischer,
J., Lockman, F.~J., \& Stemwedel, S.\ 1992, \apj, 397, 174


\bibitem[Oey \& Garc{\'{\i}}a-Segura(2004)]{Oey04} Oey,
M.~S., \& Garc{\'{\i}}a-Segura, G.\ 2004, \apj, 613, 302



\bibitem[Ogura et al.(2002)]{Ogura02} Ogura, K., Sugitani, K.,
\& Pickles, A.\ 2002, \aj, 123, 2597



\bibitem[Ogura et al.(2006)]{Ogura06} Ogura, K., Chauhan, N.,
Pandey, A.~K., Bhatt, B.~C., Ojha, D., \& Itoh, Y.\ 2006, \pasj,
in press, astro-ph/0611845



\bibitem[Oliveira et al.(2002)]{Oliviera02} Oliveira, J.~M.,
Jeffries, R.~D., Kenyon, M.~J., Thompson, S.~A., \& Naylor, T.\
2002, \aap, 382, L22



\bibitem[Oliveira et al.(2004)]{Oliviera04} Oliveira, J.~M.,
Jeffries, R.~D., \& van Loon, J.~T.\ 2004, \mnras, 347, 1327



\bibitem[Palla \& Stahler(1999)]{Palla99} Palla, F., \&
Stahler, S.~W.\ 1999, \apj, 525, 772


\bibitem[Palla \& Stahler(2000)]{Palla00} Palla, F., \&
Stahler, S.~W.\ 2000, \apj, 540, 255



\bibitem[Palla \& Stahler(2002)]{Palla02} Palla, F., \&
Stahler, S.~W.\ 2002, \apj, 581, 1194


\bibitem[Palla et al.(2005)]{Palla05} Palla, F., Randich, S.,
Flaccomio, E., \& Pallavicini, R.\ 2005, \apjl, 626, L49



\bibitem[Palla et al.(2007)]{Palla07} Palla, F., Randich, S.,
Pavlenko, Y.~V., Flaccomio, E., \& Pallavicini, R.\ 2007, \apjl,
659, L41


\bibitem[Paolillo et al.(2004)]{Paolillo04} Paolillo, M.,
Schreier, E.~J., Giacconi, R., Koekemoer, A.~M., \& Grogin, N.~A.\
2004, \apj, 611, 93


\bibitem[Porter \& Rivinius(2003)]{Porter03} Porter, J.~M., \&
Rivinius, T.\ 2003, \pasp, 115, 1153



\bibitem[Preibisch \& Zinnecker(1999)]{Preibisch99} Preibisch, T.,
\& Zinnecker, H.\ 1999, \aj, 117, 2381



\bibitem[Preibisch \& Feigelson(2005)]{Preibisch05a} Preibisch, T.,
\& Feigelson, E.~D.\ 2005a, \apjs, 160, 390



\bibitem[Preibisch et al.(2005)]{Preibisch05b} Preibisch, T., et
al.\ 2005b, \apjs, 160, 401


\bibitem[Pye \& McHardy(1983)]{Pye83} Pye, J.~P., \& McHardy,
I.~M.\ 1983, \mnras, 205, 875



\bibitem[Reipurth(1983)]{Reipurth83} Reipurth, B.\ 1983, \aap,
117, 183



\bibitem[Robin et al.(2003)]{Robin03} Robin, A.~C., Reyl{\' e},
C., Derri{\` e}re, S., \& Picaud, S.\ 2003, \aap, 409, 523



\bibitem[Sacco et al.(2006)]{Sacco06} Sacco, G.~G., Randich,
S., Franciosini, E., Pallavicini, R., \& Palla, F.\ 2006, \aap, in
press (astro-ph/0611880)



\bibitem[Santos et al.(1998)]{Santos98} Santos, N.~C., Yun,
J.~L., Santos, C.~A., \& Marreiros, R.~G.\ 1998, \aj, 116, 1376



\bibitem[Schmitt et al.(1995)]{Schmitt95} Schmitt, J.~H.~M.~M.,
Fleming, T.~A., \& Giampapa, M.~S.\ 1995, \apj, 450, 392



\bibitem[Schmitt(1997)]{Schmitt97} Schmitt, J.~H.~M.~M.\ 1997,
\aap, 318, 215



\bibitem[Siess et al.(2000)]{Siess00} Siess, L., Dufour, E., \&
Forestini, M.\ 2000, \aap, 358, 593


\bibitem[Slesnick et al.(2004)]{Slesnick04} Slesnick, C.~L.,
Hillenbrand, L.~A., \& Carpenter, J.~M.\ 2004, \apj, 610, 1045



\bibitem[Slesnick et al.(2006)]{Slesnick06} Slesnick, C.~L.,
Carpenter, J.~M., Hillenbrand, L.~A., \& Mamajek, E.~E.\ 2006,
\aj, 132, 2665



\bibitem[Smith et al.(2001)]{Smith01} Smith, R.~K., Brickhouse,
N.~S., Liedahl, D.~A., \& Raymond, J.~C.\ 2001, \apjl, 556, L91


\bibitem[Stassun et al.(2006)]{Stassun06} Stassun, K.~G., van den
Berg, M., Feigelson, E., \& Flaccomio, E.\ 2006, \apj, 649, 914



\bibitem[Steffen et al.(2004)]{Steffen04} Steffen, A.~T., Barger,
A.~J., Capak, P., Cowie, L.~L., Mushotzky, R.~F., \& Yang, Y.\
2004, \aj, 128, 1483


\bibitem[Stelzer et al.(2005)]{Stelzer05} Stelzer, B., Flaccomio,
E., Montmerle, T., Micela, G., Sciortino, S., Favata, F.,
Preibisch, T., \& Feigelson, E.~D.\ 2005, \apjs, 160, 557


\bibitem[Stelzer et al.(2006)]{Stelzer06} Stelzer, B.,
Hu{\'e}lamo, N., Micela, G., \& Hubrig, S.\ 2006, \aap, 452, 1001


\bibitem[Sugitani et al.(1991)]{Sugitani91} Sugitani, K., Fukui,
Y., \& Ogura, K.\ 1991, \apjs, 77, 59



\bibitem[Sugitani \& Ogura(1994)]{Sugitani94} Sugitani, K., \&
Ogura, K.\ 1994, \apjs, 92, 163



\bibitem[Sugitani et al.(1995)]{Sugitani95} Sugitani, K., Tamura,
M., \& Ogura, K.\ 1995, \apjl, 455, L39


\bibitem[Telleschi et al.(2006)]{Telleschi06} Telleschi, A.,
Guedel, M., Briggs, K.~R., Audard, M., \& Palla, F.\ 2006, \aap,
in press (astro-ph/0612338)


\bibitem[Thompson et al.(2004)]{Thompson04} Thompson, M.~A.,
White, G.~J., Morgan, L.~K., Miao, J., Fridlund, C.~V.~M., \&
Huldtgren-White, M.\ 2004, \aap, 414, 1017


\bibitem[Thompson et al.(2004b)]{Thompson04b} Thompson, M.~A.,
Urquhart, J.~S., \& White, G.~J.\ 2004b, \aap, 415, 627



\bibitem[Townsley et al.(2002)]{Townsley02} Townsley, L.~K.,
Broos, P.~S., Nousek, J.~A., \& Garmire, G.~P.\ 2002, Nuclear
Instruments and Methods in Physics Research A, 486, 751



\bibitem[Townsley et al.(2003)]{Townsley03} Townsley, L.~K.,
Feigelson, E.~D., Montmerle, T., Broos, P.~S., Chu, Y.-H., \&
Garmire, G.~P.\ 2003, \apj, 593, 874



\bibitem[Townsley et al.(2006)]{Townsley06} Townsley, L.~K.,
Broos, P.~S., Feigelson, E.~D., Garmire, G.~P., \& Getman, K.~V.\
2006, \aj, 131, 2164



\bibitem[Urquhart et al.(2006)]{Urquhart06} Urquhart, J.~S.,
Thompson, M.~A., Morgan, L.~K., \& White, G.~J.\ 2006, \aap, 450,
625



\bibitem[van den Heuvel(1983)]{vandenHeuvel83} van den Heuvel,
E.~P.~J.\ 1983, Accretion-Driven Stellar X-ray Sources, 303


\bibitem[van Till et al.(1975)]{vanTill75} van Till, H., Loren,
R., \& Davis, J.\ 1975, \apj, 198, 235




\bibitem[Vuong et al.(2003)]{Vuong03} Vuong, M.~H., Montmerle,
T., Grosso, N., Feigelson, E.~D., Verstraete, L., \& Ozawa, H.\
2003, \aap, 408, 581



\bibitem[Webb et al.(1999)]{Webb99} Webb, R.~A., Zuckerman,
B., Platais, I., Patience, J., White, R.~J., Schwartz, M.~J., \&
McCarthy, C.\ 1999, \apjl, 512, L63



\bibitem[Weisskopf et al.(2002)]{Weisskopf02} Weisskopf, M.~C.,
Brinkman, B., Canizares, C., Garmire, G., Murray, S., \& Van
Speybroeck, L.~P.\ 2002, \pasp, 114, 1



\bibitem[White(1993)]{White93} White, G.~J.\ 1993, \aap, 274,
L33



\bibitem[Williams et al.(1977)]{Williams77} Williams, P.~M.,
Brand, P.~W.~J.~L., Longmore, A.~J., \& Hawarden, T.~G.\ 1977,
\mnras, 180, 709


\bibitem[Wolk et al.(2005)]{Wolk05} Wolk, S.~J., Harnden,
F.~R., Jr., Flaccomio, E., Micela, G., Favata, F., Shang, H., \&
Feigelson, E.~D.\ 2005, \apjs, 160, 423



\bibitem[Yonekura et al.(1999)]{Yonekura99} Yonekura, Y.,
Hayakawa, T., Mizuno, N., Mine, Y., Mizuno, A., Ogawa, H., \&
Fukui, Y.\ 1999, \pasj, 51, 837



\bibitem[Zacharias et al.(2004a)]{Zacharias04a} Zacharias, N., Monet,
D.~G., Levine, S.~E., Urban, S.~E., Gaume, R., \& Wycoff, G.~L.\
2004a, Bulletin of the American Astronomical Society, 36, 1418



\bibitem[Zacharias et al.(2004b)]{Zacharias04b} Zacharias, N., Urban,
S.~E., Zacharias, M.~I., Wycoff, G.~L., Hall, D.~M., Monet, D.~G.,
\& Rafferty, T.~J.\ 2004b, \aj, 127, 3043


\bibitem[Ziolkowski(2002)]{Ziolkowski02} Ziolkowski, J.\ 2002,
Memorie della Societa Astronomica Italiana, 73, 1038



\end{thebibliography}
\end{document}